%% file: KrLoMu_SampledNonlinDetect2023_rev.tex
\documentclass[journal,twoside,web]{ieeecolor}
\usepackage{generic}
\usepackage{cite}
\usepackage{amsmath,amssymb,amsfonts}
\usepackage{algorithmic}
\usepackage{graphicx}
\usepackage{algorithm,algorithmic}
\usepackage{hyperref}

\usepackage{epsfig} 

\usepackage{pgfkeys}
\usepackage{xcolor}
\usepackage{pgfplots}
\usepackage{layouts}
\usepackage{tikz}
\usepackage[ngerman]{babel}

\DeclareMathOperator*{\argmax}{arg\,max}

\shorthandon{"}
\def \iIOSS/{i"~IOSS}
\def \iISS/{i"~ISS}
\shorthandoff{"}

\definecolor{new}{rgb}{0,0,0}
\definecolor{rev}{rgb}{0,0,0}   
\usepackage{interval}
\newlength\figurewidth

\hypersetup{hidelinks=true}
\usepackage{textcomp}
\def\BibTeX{{\rm B\kern-.05em{\sc i\kern-.025em b}\kern-.08em
    T\kern-.1667em\lower.7ex\hbox{E}\kern-.125emX}}
\newcommand\copyrighttext{%
	\footnotesize \copyright 2024 IEEE. Personal use of this material is permitted. Permission from IEEE must be obtained for all other uses, in any current or future media, including reprinting/republishing this material for advertising or promotional purposes, creating new collective works, for resale or redistribution to servers or lists, or reuse of any copyrighted component of this work in other works.}
\newcommand\copyrightnotice{%
	\begin{tikzpicture}[remember picture,overlay]
		\node[anchor=south,yshift=7pt] at (current page.south) {\fbox{\parbox{\dimexpr\textwidth-\fboxsep-\fboxrule\relax}{\copyrighttext}}};
	\end{tikzpicture}%
}
\begin{document}
	\newtheorem{defi}{Definition}
	\newtheorem{thm}{Theorem}
	\newtheorem{lem}{Lemma}
	\newtheorem{rem}{Remark}
	\newtheorem{ass}{Assumption}
	\newtheorem{prop}{Proposition}
	\newtheorem{cor}{Corollary}
	
\title{Sample-based nonlinear detectability for discrete-time systems}
\author{Isabelle Krauss, Victor G. Lopez, \IEEEmembership{Member, IEEE}, and Matthias A. Müller,  \IEEEmembership{Senior Member, IEEE}
\thanks{This work received funding from the European Research Council (ERC) under the European Union’s Horizon 2020 research and innovation programme (grant agreement No 948679).}
\thanks{I. Krauss, V. G. Lopez and M. A. Müller are with the Institute of Automatic Control, Leibniz University Hannover, 30167 Hannover, Germany
		{\tt\small \{krauss,lopez,mueller\}@irt.uni-hannover.de}
}
}

\maketitle
\thispagestyle{empty}
\pagestyle{empty}
\copyrightnotice

\begin{abstract}
This paper introduces two sample-based formulations of incremental input/output-to-state stability (\iIOSS/), a suitable detectability notion for general nonlinear systems. In this work we consider the case of limited output information, i.e., measurements are only infrequently and/or irregularly available. The output-dependent term of the sample-based \iIOSS/ bound is properly  modified to yield a characterization for detectability in  presence of incomplete output sequences. We provide both a non-time-discounted and a time-discounted formulation of sample-based \iIOSS/. Furthermore, conditions for an \iIOSS/ system to be also sample-based \iIOSS/ are given and the relation between the two formulations of sample-based \iIOSS/ is shown.
\end{abstract}


\section{Introduction}
\label{sec:introduction}
\IEEEPARstart{I}{n} many engineering applications like state-feedback control and system monitoring, an estimate of the internal state of the system is required. There are applications where only infrequent and/or irregular output measurements of a system are available, e.g., due to the characteristics of the application that make it impractical or impossible to measure the output continuously or at every time instant. Such cases of limited output information can arise for example in the biomedical field, when analyzing blood samples of a patient, which are only taken infrequently. These measurements must then be used, for instance, to determine certain hormone concentrations in order to detect disorders of the hypothalamic–pituitary–thyroid
axis \cite{Die16} and to devise suitable medication strategies, compare,
e.g., \cite{Bru21, Wol22}. Furthermore, event-triggered sampling of networked control systems can generate such irregular sampling sequences to reduce computation and communication loads~\cite{Ge20, Mis15,Pen18}. Under such circumstances suitable state estimators that can handle this limited output information to still recover the internal state are needed.
\par To design suitable state estimators, we are therefore interested in sample-based observability or detectability conditions that take this limited output information into account. In \cite{Wan11} and \cite{Zen16} observability of linear continuous-time systems under irregular sampling is studied. These papers provide a lower bound on the number of arbitrary samples in a fixed time interval to guarantee that the internal state can be reconstructed by the irregularly sampled output. In \cite{Kra22} sample-based observability of linear discrete-time systems is investigated. There, a sample-based version  of the observability rank condition is proposed that directly takes the irregular and/or infrequent measurement instances into account and conditions on the sampling schemes are provided to render the system sample-based observable. 
\par In recent years, optimization-based state estimation techniques such as moving horizon estimation (MHE) have gained an increasing amount of attention. This is mostly due to the fact that strong stability guarantees can be shown for general nonlinear systems, and  that additionally known (physical) constraints on states and/or disturbances (such as nonnegativity constraints for concentrations of certain chemical substances) can be  incorporated in the estimation problem. In the context of nonlinear MHE, the concept of incremental input/output-to-state stability (\iIOSS/) has turned out to be a suitable notion of nonlinear detectability that by now is the standard condition used to achieve robust stability results 
\cite{All21,Hu17,Ji16,Mul17,Raw12,Sch23}. 
The notion of \iIOSS/ was first introduced in \cite{Son95} where it is also shown that \iIOSS/ is a necessary condition for the existence of a full-order state observer.
In  \cite{All20} and \cite{Knu20} time-discounted formulations of \iIOSS/ are proposed, where the \iIOSS/ bound is characterized in terms of time-discounted inputs and outputs instead of just the largest inputs and outputs.
\par In this paper, we focus on a sample-based version of \iIOSS/ that takes only an infrequent and/or irregular measurement sequence into account, in order to address the sample-based observability/detectability problem for general discrete-time nonlinear systems.
\par In particular, our contributions are as follows. We introduce the notion of sample-based \iIOSS/ as well as a time-discounted version of it. Here, the irregular or infrequent measurement sequence is directly considered in the output-dependent terms of the \iIOSS/ bounds. We provide a sufficient condition for a discrete-time \iIOSS/ system to be both sample-based \iIOSS/ and sample-based time-discounted \iIOSS/. This condition basically demands that after some finite time the magnitudes of the missing outputs are bounded  by the previous inputs and by the infrequently measured outputs. Furthermore, it is investigated when such a condition is also necessary for the system to be sample-based (time-discounted) \iIOSS/. The results of this paper pave the way to design (optimization-based) state estimators for nonlinear systems in case that only infrequent and/or irregular measurements are available.
\par The rest of the paper is organized as follows. In Section~\ref{sec:Pre}, we state some technical definitions. In Section~\ref{sec:sbiISS} we propose the notion of sample-based \iIOSS/. Conditions for an \iIOSS/ system to be sample-based \iIOSS/ are also presented. Section~\ref{sec:sbtdiISS} introduces a time-discounted version of sample-based \iIOSS/ and explores conditions for a system to be sample-based time-discounted \iIOSS/. Furthermore, the relation between the two proposed characterizations of sample-based detectability is studied. Finally, Section \ref{sec:con} concludes the paper.

\section{Preliminaries and Setup} 
\label{sec:Pre}
Consider the following nonlinear discrete-time system
\begin{align}
	\begin{aligned}
		x(t+1)&=f(x(t),w(t)) \\
		y(t)&=h(x(t))\\
	\end{aligned}
	\label{eq:sys}
\end{align}
in which $x \in \mathbb{R}^n$ is the system state, 
${w} \in \mathbb{R}^q $ is the (disturbance) input, and $y \in \mathbb{R}^p$ is the system output. 
We denote all nonnegative real numbers and all nonnegative integers by $\mathbb{R}_{\geq0} $ and $\mathbb{I}_{\geq 0}$, respectively, and all positive real numbers and integers by $\mathbb{R}_{>0}$ and $\mathbb{I}_{>0}$. The symbol $\oplus$ denotes the maximum operator according to \begin{align}
	a\oplus b:=\max\{a,b\}.
\end{align}
The Euclidean norm is denoted by\footnote{We note that any other vector norm in $\mathbb{R}^n$ can also be used throughout the document.} $|\cdot|$ and the bold symbol $\boldsymbol{w}$ refers to a sequence of the vector-valued variable ${w\in\mathbb{R}^q}, \boldsymbol{w}= \{w(0),w(1),\ldots\}$. The notation $(\mathbb{R}^q)^{\mathbb{I}_{\geq0}}$ denotes the set of all sequences $\boldsymbol{w}$ with infinite length. The notation $x(t;x_{0},\boldsymbol{w})$ denotes the solution to system (\ref{eq:sys}) at time $t$ for initial condition $x_0$ and disturbance sequence $\boldsymbol{w}$.
Since we work  with incremental stability properties, we use the following abbreviated notation 
\begin{align}
	\begin{aligned}
		\Delta x(t)&:=x(t;x_{01},\boldsymbol{w_1})-x(t;x_{02},\boldsymbol{w_2})	\\ \Delta x_0&:=x_{01}-x_{02}\\
		\Delta w(t)&:=w_1(t)-w_2(t)\\ \Delta \boldsymbol{w}&:=\boldsymbol{w_1}-\boldsymbol{w_2} 
	\end{aligned}
\end{align}
with $x_{01},x_{02}$ and \textcolor{rev}{$\boldsymbol{w_1}= \{w_1(0),w_1(1),\ldots\}, \boldsymbol{w_2}= \{w_2(0),w_2(1),\ldots\}$} being two initial states and two disturbance sequences, respectively.
\textcolor{rev}{Additionally, $|\Delta \boldsymbol{w}|$ refers to the Euclidean norm of the concatenation of the elements of the sequence $\Delta \boldsymbol{w}$.}
With a slight abuse of notation, we also use the following expression
\begin{align}
	\begin{aligned}
		\Delta h(\Delta x(t))&:=h(x(t; x_{01},\boldsymbol{w_1}))-h(x(t; x_{02},\boldsymbol{w_2})).
	\end{aligned}
\end{align}
\textcolor{rev}{The detectability conditions considered in this paper employ comparison functions according the following definition.}
\begin{defi}[Comparison functions \cite{Kel14}.]
	A function $\phi: \mathbb{R}_{\geq0} \rightarrow  \mathbb{R}_{\geq 0}$ is called a $\mathcal{K}$-function if it is continuous, strictly increasing and  $\phi(0)=0$ . 	A function $\phi: \mathbb{R}_{\geq0} \rightarrow  \mathbb{R}_{\geq 0}$ is called a $\mathcal{K}_{\infty}$-function if $\phi \in \mathcal{K}$ and,  in addition, $\lim_{s \rightarrow \infty}\phi(s)=\infty$. A function $\sigma: \mathbb{I}_{\geq0} \rightarrow  \mathbb{R}_{\geq 0}$ is called an $\mathcal{L}$-function if it is continuous, \textcolor{rev}{strictly decreasing} and $\lim_{t\rightarrow\infty}\sigma(t)=0$. A function $\beta: \mathbb{R}_{\geq 0}\times \mathbb{I}_{\geq0} \rightarrow  \mathbb{R}_{\geq 0}$ is called a $\mathcal{KL}$-function if $\beta(\cdot,t)\in \mathcal{K}$ for each fixed $t\in \mathbb{I}_{\geq0}$, and  $\beta(s,\cdot)\in \mathcal{L}$ for each fixed $s \in \mathbb{R}_{\geq 0}$.\\
	\label{def:comp}
\end{defi}
\par Furthermore, we consider in the context of sample-based detectability systems with $\mathcal{K}$- continuous output functions.\\
\textcolor{rev}{
\begin{defi}[$\mathcal{K}$-continuity \cite{Raw17}.]
	A function $g: \mathbb{X}\times \mathbb{V} \rightarrow \mathbb{Y}$, where $\mathbb{X},\mathbb{V} ,\mathbb{Y}$  are subspaces of Euclidean spaces,  is $\mathcal{K}$-continuous if there exists a function $\phi\in \mathcal{K}$ such that for all $x_1, x_2 \in \mathbb{X}$ and  $v_1, v_2 \in \mathbb{V}$
	\begin{align}
		|g(x_1,v_1)-g(x_2,v_2)|\leq \phi(|x_1-x_2|+|v_1-v_2|).
	\end{align}
	\label{def:cont}
\end{defi}
}
\begin{rem}
	Note that all the analysis and results that are presented in the rest of the paper also hold if we consider in  system (\ref{eq:sys}) the system state $x \in \mathbb{X} \subset \mathbb{R}^n $ and the input ${w} \in \mathbb{W} \subset \mathbb{R}^q$. With corresponding modifications, e.g., formulating a detectability condition on these subsets of $\mathbb{R}^n$ and $\mathbb{R}^q$, all results can be adapted to such a setting.
\end{rem}
\section{Sample-based \iIOSS/} \label{sec:sbiISS}
An appropriate detectability condition for nonlinear systems, that - as discussed in Section \ref{sec:introduction} - is particularly suitable in the context of moving horizon estimation, is incremental input/output-to-state stability (\iIOSS/).\\
\begin{defi}[\iIOSS/ \cite{Son95}] The system (\ref{eq:sys}) is \iIOSS/ if there exist functions $\beta \in \mathcal{KL}$ and 
	$\gamma_1,\gamma_2 \in  \mathcal{K}$  such that for any two initial
	conditions $x_{01}$, $x_{02}$ and any two input (disturbance) trajectories $\boldsymbol{w_1},\boldsymbol{w_2}$ the following\footnote{\textcolor{rev}{Throughout this paper we use the convention $\sup_{0 \leq j <0} g(j) := 0$.}} holds for all $t \geq 0$
	\begin{align}
		\begin{aligned}
			|\Delta x(t)|
			&\leq \beta(| \Delta x_0
			|,t	)\oplus\gamma_1(\sup_{0\leq \tau < t} | \Delta w(\tau)
			|)\\
			&\oplus\gamma_2(\sup_{0\leq \tau < t} |\Delta h(\Delta x(\tau))|).
			\label{eq:iIOSS}
		\end{aligned}
	\end{align}
	\label{def:iioss}
\end{defi}
\par If there exist functions $\beta$ and $\gamma_1$ such that the left-hand side  of (\ref{eq:iIOSS}) is bounded by the first two terms of the right-hand side, then the system is incrementally input-to-state stable (\iISS/) \cite{Ang02}, i.e., no output information is needed to bound $|\Delta x|$ and two solutions  will always converge to each other in case the same input is applied ($\Delta w \equiv 0$).
\par \textcolor{rev}{We note that a system being \iIOSS/ is equivalent to the system admitting a so-called \iIOSS/ Lyapunov function \cite{All20}.}
Recently, a systematic procedure to compute \iIOSS/ Lyapunov functions for classes of nonlinear systems (and hence to establish that a system is \iIOSS/) has been proposed in \cite[Sec. IV]{Sch23}.

In the rest of this section, we first introduce the notion of sample-based incremental input/output-to state stability. We then study conditions for an \iIOSS/ system to be sample-based \iIOSS/ as well. Here the following assumption on the system is made. \\
\begin{ass}
	The output function $h$ in (\ref{eq:sys}) is $\mathcal{K}$-continuous, meaning that there exists a function  $\alpha_h \in \mathcal{K}$ such that for all $x_1,x_2\in \mathbb{R}^n$	
	\begin{align}
		|h(x_1)-h(x_2)| \leq \alpha_h(|x_1-x_2|).
		\label{eq:Kcont}
	\end{align}
	\label{ass:h}
\end{ass} 
\par Before introducing the sample-based detectability condition, we define a set of sampling instances $K$.\\
\begin{defi}[Sampling set $K$] 
	Consider an infinitely long sequence $\textcolor{rev}{D}=\{\delta_1,\delta_2,\ldots\}$ with $\delta_i \in \mathbb{I}_{\geq 0}, \ i\in \mathbb{I}_{> 0}$ and  $\max_i \delta_i \leq \delta_{max}\in \mathbb{I}_{\geq 0}$. 
	The set $K_i=\{t_1^i,t_2^i,\ldots\}$ refers to an infinite set of time instances defined as
	\begin{align}
		\begin{aligned}
			t_1^i=&\delta_i\\
			t_2^i=&t_1^i+\delta_{i+1}\\ 
			&\vdots\\
			t_j^i=&t_{j-1}^i+\delta_{i+j-1}  \\
				&\vdots
		\end{aligned}                    
	\end{align} 
The set $K$ then refers to a set of sets containing all $K_i$, $ i\in\mathbb{I}_{> 0}$.
	\label{def:K}
\end{defi}
\par  Before explaining Definition \ref{def:K}, we define a sample-based detectability condition.\\ 
\begin{defi}[Sample-based \iIOSS/]
	Consider some set $K$ according to Definition~\ref{def:K}. The system (\ref{eq:sys}) is sample-based \iIOSS/ with respect to $K$ if there exist functions $\bar{\beta} \in \mathcal{KL}$ and $\bar{\gamma}_1,\bar{\gamma}_2 \in \mathcal{ K}$  such that for any two initial conditions $x_{01}$, $x_{02}$ and any two input (disturbance) trajectories $\boldsymbol{w_1},\boldsymbol{w_2}$ the following holds  for any $K_i  \in K$ and all $t \geq 0$ 
	\begin{align}
		\begin{aligned}
			|
			\Delta x(t)
			|
			&\leq \bar{\beta}(| \Delta x_0
			|,t	)\oplus\bar{\gamma}_1(\sup_{0\leq \tau < t} | \Delta w(\tau)
			|)\\
			&\oplus\bar{\gamma}_2(\sup_{ \tau \in K_i, \tau < t} | \Delta h(\Delta x(\tau))|).
			\label{eq:s-iIOSS}
		\end{aligned}
	\end{align}
\label{def:sbiIOSS}
\end{defi}
\par The reason for defining the sampling set $K$  as a set of sets, rather than as a set containing one specific sequence of time instances is the following. It is of our interest to analyze conditions for a sampling scheme to preserve the detectability of a system. A sampling scheme that keeps the system detectable must do so uniformly in time. This means that the state of the system can be reconstructed regardless of the time at which we start using the measured information. For example, a sampling scheme of the form ''take at least $N$ measurements every $T$ time instances'' satisfies Definition~\ref{def:K}. On the contrary, a sampling scheme that stops taking measurements after a certain time $T$, is discarded by Definition~\ref{def:K}. It is important to highlight the fact that the definition of the sampling set $K$ does not imply periodicity of the sampling instances. For further insight, note also the following example that presents the case where a sampling set $K_1$ renders the system sample-based \iIOSS/, however not all $K_i$ as in Definition \ref{def:K} guarantee sample-based \iIOSS/ as in Definition \ref{def:sbiIOSS}. Here, consider an observable single-output linear system with a complex conjugate eigenvalue pair on the unit circle, e.g., a system with the following system dynamics
\begin{align}
	\boldsymbol{x}(t+1)=\begin{pmatrix} \frac{1}{\sqrt{2}}& \frac{1}{\sqrt{2}}\\ -\frac{1}{\sqrt{2}}& \frac{1}{\sqrt{2}}\end{pmatrix} \boldsymbol{x}(t).
	\label{eq:pathperiodex}
\end{align} 
Complex eigenvalues can  cause the system to have a pathological sampling period $p$, meaning that for some initial conditions, taking a sample every $p$ time instances yields zero difference in the measured outputs, i.e., $\Delta h(\Delta x(jp))=0$ for all $j \in \mathbb{I}_{\geq 0}$. For the exemplary dynamics in (\ref{eq:pathperiodex}), $p$ takes the value $4$. For more details on pathological  sampling periods see \cite{Kra22}. Since the system is marginally stable, selecting $\textcolor{rev}{D}=\{\delta_1, \delta_2, \ldots,\delta_r, p,p,\ldots\}$ where $\delta_i \neq j p$ for $i \leq r$ and $\delta_i = p$ for all $i > r$, 
 can guarantee  by the first $r$ measurements sample-based \iIOSS/ of the system with respect to $K_1$. But, due to the pathological period the system is not sample-based \iIOSS/ with respect to $K_{r+1}$. This example illustrates that there exist cases where sample-based \iIOSS/ with respect to some set $K_1$ does not imply sample-based \iIOSS/ with respect to all $K_i$ with $i>1$. However, even though $K_1$ satisfies the sample-based \iIOSS/ condition, i.e., condition (\ref{eq:s-iIOSS}), such a sampling scheme would not be favorable in context of state estimation since after a short time no further meaningful measurements may be obtained, opposite to using a sampling sequence that renders the system sample-based \iIOSS/ with respect to $K$ as defined in Definition~\ref{def:K}. Later, in Section~\ref{sec:sbtdiISS}, this definition of $K$ will also be useful to address the time-discounted detectability case.
\subsection{Sufficient condition for sample-based i-IOSS}
The following theorem gives a sufficient condition for an \iIOSS/ system to be sample-based \iIOSS/.\\
\begin{thm}
	Let system (\ref{eq:sys}) be \iIOSS/ and let Assumption~\ref{ass:h} hold.  Moreover, consider  some set $K$ according to Definition~\ref{def:K}. Then, the system is sample-based \iIOSS/ with respect to $K$ if there exist functions $\gamma_w, \gamma_h \in \mathcal{K}$ and a finite time $t^*$ such that for any two initial conditions $x_{01}$, $x_{02}$ and two input (disturbance) trajectories $\boldsymbol{w_1},\boldsymbol{w_2}$ the following holds  
	\begin{align}
			\begin{aligned}
		&| \Delta h(\Delta x(t))|\leq \gamma_h( \sup_{ \tau \in K_i, \tau < t} | \Delta h(\Delta x(\tau))|)\\ &\oplus\gamma_w(\sup_{0\leq \tau < t} | \Delta w(\tau)
		|), \quad \forall t\geq t^*, \ \forall K_i \in K.
			\end{aligned}
		\label{eq:cond_sampling}
	\end{align}
	\label{thm1}
\end{thm}
\begin{proof}
	Consider some function $\tilde{h}_t$ defined as
	\begin{align}
		\tilde{h}_t(x_0,\boldsymbol{w}):=h(x(t;x_0,\boldsymbol{w})).
	\end{align}
Since the system is \iIOSS/, we can use (\ref{eq:iIOSS}) for $t=1$ together with $\mathcal{K}$-continuity of $h$ to conclude that $f$ is $\mathcal{K}$-continuous as well.
	Recursively applying $\mathcal{K}$-continuity of $f$ together with $\mathcal{K}$-continuity of $h$ shows that $\tilde{h}_t$ is also $\mathcal{K}$-continuous. Hence, for each $t\in [0,t^*)$, there exists a function $\alpha_{h_t}\in \mathcal{K}$ such that
	\begin{align} 
		|\tilde{h}_t(x_{01},\boldsymbol{w_1})-\tilde{h}_t(x_{02},\boldsymbol{w_2})|\leq \alpha_{h_t}(|\Delta x_0|+|\Delta \boldsymbol{w}|).
	\end{align}  
	Now, given the finite time instance $t^*$, 	\textcolor{rev}{define a function} $\tilde{\alpha}_h$ such that 
	\begin{align}
		\tilde{\alpha}_{h}\geq \alpha_{h_t}, \quad \forall [0,t^*].
		\label{eq:Lh}
	\end{align}
	Then, it holds that
	\begin{align}
		\begin{aligned}
		|\tilde{h}_{t}(x_{01},{\boldsymbol{w_1}})-\tilde{h}_t(x_{02},{\boldsymbol{w_2}})|\leq \tilde{\alpha}_h (|\Delta x_0|+|\Delta \boldsymbol{w}|), \\ \forall t \in  [0,t^*]. 
		\label{eq:K_cont_h}
		\end{aligned}
	\end{align} 
	Now we can formulate an upper bound for the output-dependent term  $\gamma_2(\sup_{0\leq \tau < t} | \Delta h(\Delta x(\tau))|)$ of  the \iIOSS/ bound~(\ref{eq:iIOSS}) for $t\leq t^*$.  By condition (\ref{eq:K_cont_h}) we obtain
	\begin{align}
		\begin{aligned}
		\gamma_2(\sup_{0\leq \tau < t} | \Delta h(\Delta x(\tau))|) \leq\gamma_2(\tilde{\alpha}_h(|\Delta x_0|+|\Delta \boldsymbol{w}|))\\\leq \gamma_2(\tilde{\alpha}_h(2|\Delta x_0|))\oplus\gamma_2(\tilde{\alpha}_h(2|\Delta \boldsymbol{w}|)).
		\label{eq:gamma2_l}
			\end{aligned}
	\end{align}
    Inserting (\ref{eq:gamma2_l}) in (\ref{eq:iIOSS}) yields for all $[0,t^*]$ 
    the following upper bound for $|\Delta x(t)|$
	\begin{align}
		\begin{aligned}
			|
			\Delta x(t)
			|
			&\leq \beta(| \Delta x_0
			|,t	)\oplus\gamma_1(\sup_{0\leq \tau < t} | \Delta w(\tau)
			|)\\
			&\oplus\gamma_2(\tilde{\alpha}_h(2|\Delta x_0|))\oplus\gamma_2(\tilde{\alpha}_h(2|\Delta \boldsymbol{w}|)).
			\label{eq:iIOSS_beta}
		\end{aligned}
	\end{align}
\textcolor{rev}{
	We now consider the case\footnote{\textcolor{rev}{Note that the time instant $t^*$ is considered in both cases. This is not  necessary, but it allows for the simplification of certain steps in the proof.}} $t \geq t^*$. From (\ref{eq:iIOSS}) and taking $(x(t^*;x_{01},\boldsymbol{w_1}), x(t^*;x_{02},\boldsymbol{w_2}))$ as the initial conditions of some trajectories, we can write the following
\begin{align}
		\begin{aligned}
			|\Delta x(t)|
			&\leq \beta(| \Delta x(t^*)
			|,t-t^*	)\oplus\gamma_1(\sup_{t^*\leq \tau < t} | \Delta w(\tau)
			|)\\
			&\oplus\gamma_2(\sup_{t^*\leq \tau < t} |\Delta h(\Delta x(\tau))|).
			\label{eq:iIOSStstar}
		\end{aligned}
\end{align}
By condition (\ref{eq:cond_sampling}), we can upper bound the output-dependent term of (\ref{eq:iIOSStstar}) as follows  
	\begin{align}
				\begin{aligned}
				\gamma_2(\sup_{t^*\leq \tau < t} | \Delta h(\Delta x(\tau))|) \leq \gamma_2\Bigl(\gamma_w(\sup_{0\leq \tau < t} | \Delta w(\tau)
				|)\\ \oplus \gamma_h( \sup_{ \tau \in K_i, \tau < t} | \Delta h(\Delta x(\tau))|) \Bigr). \label{eq:gamma2_g}
					\end{aligned}
		\end{align}
In the following, it will be convenient to find an additional upper bound for $|\Delta x(t^*)|$. 
Using again the i-IOSS bound (\ref{eq:iIOSS}) to upper bound $|\Delta x(t^*)|$  yields
	\begin{align}
			\begin{aligned}
				|\Delta x(t^*)|
				&\leq \beta(| \Delta x_0
				|,t^*	)\oplus\gamma_1(\sup_{0\leq \tau < t^*} | \Delta w(\tau)
				|)\\
				&\oplus\gamma_2(\sup_{0\leq \tau < t^*} |\Delta h(\Delta x(\tau))|).
				\label{eq:iIOSStstar2}
			\end{aligned}
		\end{align}
Substituting (\ref{eq:gamma2_l}), (\ref{eq:gamma2_g}) and (\ref{eq:iIOSStstar2}) in (\ref{eq:iIOSStstar})  we finally obtain the following upper bound for $|\Delta x(t)|$ for  $t\geq t^*$
	\begin{align}
		\begin{aligned}
		|
		\Delta x(t)
		|
		&\leq \beta(\beta(| \Delta x_0
		|,t^*),t-t^*	)\\&\oplus \beta(\gamma_1(\sup_{0\leq \tau < t^*} | \Delta w(\tau)
		|),t-t^*	)\\&\oplus \beta(\gamma_2(\tilde{\alpha}_h(2|\Delta x_0|)),t-t^*	)\\&\oplus \beta(\gamma_2(\tilde{\alpha}_h(2|\Delta \boldsymbol{w}|)),t-t^*	)\\&
		\oplus\gamma_1(\sup_{t^*\leq \tau < t} | \Delta w(\tau)
		|)\\
		&\oplus	\gamma_2(\gamma_w(\sup_{0\leq \tau < t} | \Delta w(\tau)
		|) )\\&\oplus\gamma_2(\gamma_h( \sup_{ \tau \in K_i, \tau < t} | \Delta h(\Delta x(\tau))|)).
		\label{eq:iIOSS_beta2}
	\end{aligned}
	\end{align}
}

	\par \textcolor{rev}{Next, we derive expressions for the functions $\bar{\beta},\bar{\gamma}_1,\bar{\gamma}_2$ satisfying the sample-based \iIOSS/ bound in (\ref{eq:s-iIOSS}). By (\ref{eq:iIOSS_beta}) and (\ref{eq:iIOSS_beta2}) we obtain the following lower bound for $\bar{\beta}$
	\begin{align}
	&\bar{\beta}(|\Delta x_0|,t)  \label{eq:betabar}\\ 
	&\geq\begin{cases}
			\beta(|\Delta x_0|,t)\oplus \gamma_2(\tilde{\alpha}_h(2|\Delta x_0|)),&0\leq t<t^* \nonumber\\
			\beta(\beta(| \Delta x_0
			|,t^*)\oplus \gamma_2(\tilde{\alpha}_h(2|\Delta x_0|)),t-t^*	),&t\geq t^*. \nonumber
		\end{cases} 
	\end{align}
}
\textcolor{rev}{
	To satisfy (\ref{eq:betabar}), we propose the following expression of $\bar{\beta}$ 
	\begin{align}
			\begin{aligned}
		\bar{\beta}(|\Delta x_0|,t)&=\beta (|\Delta x_0|,t)\oplus \gamma_2(\tilde{\alpha}_h(2|\Delta x_0|))e^{t^*-t}\\&\oplus \	\hat{\beta}(| \Delta x_0
		|,t ).
			\end{aligned} 
		\label{eq:betabar2}
	\end{align}
	where 
	\begin{align}
	&	\hat{\beta}(| \Delta x_0
		|,t ):=\\&\begin{cases}
		\beta(\beta(| \Delta x_0
		|,t^*)\oplus \gamma_2(\tilde{\alpha}_h(2|\Delta x_0|)),0	)e^{t^*-t}, &0\leq t<t^* \nonumber\\
		\beta(\beta(| \Delta x_0
		|,t^*)\oplus \gamma_2(\tilde{\alpha}_h(2|\Delta x_0|)),t-t^*	),& t\geq t^*. \nonumber
		\end{cases}
	\end{align}
It can be readily observed that all terms on the right-hand side of (\ref{eq:betabar2})  are $\mathcal{KL}$-functions, and therefore so is $\bar{\beta}$.
	Clearly, the second inequality in (\ref{eq:betabar}) is always satisfied by the selection in (\ref{eq:betabar2}). 
	Moreover, for $t<t^*$,  \textcolor{rev}{$e^{t^*-t}> 1$}. Hence,
	\begin{align}
		\bar{\beta}(|\Delta x_0|,t)\geq \beta(|\Delta x_0|,t)\oplus\gamma_2(\tilde{\alpha}_h(2|\Delta x_0|)), \ \forall t<t^*,
	\end{align}
	Thus, the  proposed $\bar{\beta}$  in (\ref{eq:betabar2}) satisfies the inequalities in (\ref{eq:betabar}).}
	\par \textcolor{rev}{Considering all input-dependent terms in (\ref{eq:iIOSS_beta}) and (\ref{eq:iIOSS_beta2}), a lower bound for $\bar{\gamma}_1$ guaranteeing to make the sample-based \iIOSS/ bound hold can be formulated. Note that, since $\beta$ satisfies the i-IOSS bound, $\beta(s,0)\geq s$ for all $s\in \mathbb{R}_{\geq0}$. The lower bound is then given as follows
	\begin{align}
	\begin{aligned}
\bar{\gamma}_1(\sup_{0\leq \tau < t} | \Delta w(\tau)|)&\geq
		\beta(\gamma_1(\sup_{0\leq \tau < t} | \Delta w(\tau)|),0)\\&
		\oplus\beta(\gamma_2(\tilde{\alpha}_h( 2t^*\sup_{0\leq \tau < t} | \Delta w(\tau)|)),0)\\
		& \oplus\gamma_2(\gamma_w(\sup_{0\leq \tau < t} | \Delta w(\tau)|)).
	\label{eq:gamma_1}
\end{aligned}
	\end{align}
}
	This can be seen since, for the case of $0\leq t<t^*$, we can bound $|\Delta \boldsymbol{w}|$ by
	\begin{align}
		\begin{aligned}
			|\Delta \boldsymbol{w}|&\leq \sum_{\tau=0}^{t-1}| \Delta w(\tau)|\leq t\sup_{0\leq \tau < t} | \Delta w(\tau)|\\
			& \leq t^*\sup_{0\leq \tau < t} | \Delta w(\tau)|, \quad \forall t< t^*.
		\end{aligned}
	\end{align}
	Therefore,
	\begin{align}
		\gamma_2(\tilde{\alpha}_h(2|\Delta \boldsymbol{w}|))\leq \gamma_2(\tilde{\alpha}_h( 2t^*\sup_{0\leq \tau < t} | \Delta w(\tau)|)).
	\end{align} 
	\par  By the last term of the upper bound in (\ref{eq:iIOSS_beta2}), $\bar{\gamma}_2$ can be chosen as  
	\begin{align}
		 \bar{\gamma}_2=  \gamma_2\circ\gamma_h.
	\end{align}
	In summary, with these choices of $\bar{\beta}, \bar{\gamma}_1,$ and $\bar{\gamma}_2$, (\ref{eq:s-iIOSS}) holds and hence the system is sample-based i-IOSS.
\end{proof}
\begin{figure}[!t]
	\setlength\figurewidth{0.8\columnwidth} 
	\centerline{\input{figures/tp_mod}}
	\caption{Output difference $\Delta y(t)$ of two separating trajectories and irregular, infrequent sampled $\Delta y(t)$ of the same trajectories.}
	\label{fig1}
\end{figure}
\par Intuitively, condition (\ref{eq:cond_sampling}) implies that increments in the output of the system are noticeable in the (few) available measurements. To clarify this idea, consider the example in Figure \ref{fig1} which 
shows the output difference $\Delta y(t)$ of two trajectories of an unstable system without any  inputs $w$, as well as an irregular measurement sequence of the same trajectories. Notice that the maximum of the true difference $\Delta y(t)$ increases much faster than the maximum of the irregularly sampled  $\Delta y(t)$. Thus, the non-measured outputs cannot be bounded by the measured outputs as in (\ref{eq:cond_sampling}). Hence, the sampling scheme in Figure \ref{fig1} does not guarantee sample-based detectability by Theorem~\ref{thm1}.
\subsection{Necessary condition for sample-based i-IOSS}
In general, the condition in Theorem~\ref{thm1} is not necessary for a system to be sample-based \iIOSS/. This can be easily seen
by considering an \iISS/ system. An \iISS/ system is sample-based \iIOSS/ with respect to any set $K$, even if $K$ is empty. However, it is clear that for an empty sampling set $K$, (\ref{eq:cond_sampling}) may not hold for all $x_{01},x_{02},\boldsymbol{w_1}, \boldsymbol{w_2}$. 
In the remainder of this section, we analyze under which conditions (\ref{eq:cond_sampling}) is necessary for sample-based  \iIOSS/. First, notice that a system that is not \iISS/ may still have what we call pairs of \iISS/ trajectories.\\
\begin{defi}[Pair of \iISS/ trajectories \textcolor{rev}{w.r.t. $\{\beta,\gamma_1,\sigma_1\}$}]
	Let the system be \iIOSS/ with $\beta\in \mathcal{KL}$, $\gamma_1, \gamma_2 \in \mathcal{K}$ satisfying the \iIOSS/ bound in (\ref{eq:iIOSS}) and consider some function $\sigma_1 \in \mathcal{L}$.
	Two state trajectories defined by their initial states $x_{01}, x_{02}$ and input sequences $\boldsymbol{w_1}, \boldsymbol{w_2}$ are a pair of \iISS/ trajectories if they are bounded as follows 
	\begin{align}
		\begin{aligned}
		|\Delta x(t)|
		&\leq \beta(| \Delta x_0
		|,t	)\\&\oplus \sup_{0\leq \tau < t}\gamma_1(| \Delta w(\tau)
		|) \sigma_1(t-\tau-1),\quad \forall t\geq 0. \label{eq:Omega}
		\end{aligned}
	\end{align}
Let $\Lambda_{\iISS/}$ be defined as the set that contains all pairs of initial states and input sequences generating a pair of \iISS/ trajectories \textcolor{rev}{w.r.t. $\{\beta,\gamma_1,\sigma_1\}$}, i.e., (\ref{eq:Omega}) holds for all $(x_{01}, x_{02},\boldsymbol{w_1}, \boldsymbol{w_2}) \in \Lambda_{\iISS/}$.
	\label{def:iisstraj}
\end{defi}

\par Pairs of i-ISS trajectories can be interpreted as trajectories that show properties like those of an i-ISS system. Namely, they can be bounded in  terms of their initial states and inputs, and furthermore converge to each other for converging inputs, i.e, $x(t;x_{01},\boldsymbol{w_1}) \rightarrow x(t;x_{02},\boldsymbol{w_2})$ if $w_1(t) \rightarrow w_2(t)$ as $t \rightarrow \infty$. The convergence of the trajectories when $\Delta w(t)\rightarrow 0$ as $t \rightarrow \infty$ is guaranteed by using $\sigma_1\in \mathcal{L}$ in (\ref{eq:Omega}), and hence discounting the influence of inputs in the past. This is related to time-discounted expressions for \iIOSS/, which are known to exist for every system that satisfies (\ref{eq:iIOSS}), i.e., is \iIOSS/ according to Definition \ref{def:iioss} \cite[Proposition 4]{All20}.
Time-discounting will also be further discussed in Section \ref{sec:sbtdiISS}.

\par Even if system (\ref{eq:sys}) is not \iISS/, the presence of pairs of \iISS/ trajectories may allow (\ref{eq:sys}) to be sample-based \iIOSS/ without the need for (\ref{eq:cond_sampling}) to hold for all initial conditions as required in Theorem~\ref{thm1}. 
 In this case, it is only needed that for each $(x_{01}, x_{02},\boldsymbol{w_1}, \boldsymbol{w_2}) \notin \Lambda_{\iISS/}$, $|\Delta h(\Delta x(t))|$ is bounded by the previous inputs and measured outputs from some finite time $t^*$ on. 
However, to guarantee that there exist functions $\gamma_w, \gamma_h \in \mathcal{K}$ and a finite time $t^*$  such that (\ref{eq:cond_sampling}) is satisfied for all \mbox{$(x_{01}, x_{02},\boldsymbol{w_1}, \boldsymbol{w_2}) \notin \Lambda_{\iISS/}$}, in general we need to restrict the disturbance sequences to a set $ W\subseteq  (\mathbb{R}^q)^{\mathbb{I}_{\geq0}}$.
To this end, we define the following set.\\
\begin{defi}[$\Psi$]
	Let system (\ref{eq:sys}) be \iIOSS/ with $\beta\in \mathcal{KL}$, $\gamma_1, \gamma_2 \in \mathcal{K}$ satisfying the \iIOSS/ bound in (\ref{eq:iIOSS}) and 
	consider an input set $W\subseteq (\mathbb{R}^q)^{\mathbb{I}_{\geq0}}$. The set $\Psi$ refers to all $(x_{01},x_{02},\boldsymbol{w_1},\boldsymbol{w_2})$  with $\boldsymbol{w_1},\boldsymbol{w_2}\in W$ 
	for which (\ref{eq:Omega}) does not hold.\\
	\label{def:Psi}
\end{defi}
\par This definition means that the set $\Psi$ contains all pairs of initial conditions and inputs (the latter contained in the set $W$) that do not result in pairs of i-ISS trajectories as defined in Definition \ref{def:iisstraj}.
\par Notice that the sufficient condition in Theorem \ref{thm1} to guarantee sample-based \iIOSS/ is that (\ref{eq:cond_sampling}) holds for all initial conditions and all input sequences. Definition \ref{def:Psi} straightforwardly allows us to relax this condition, since it is enough for (\ref{eq:cond_sampling}) to hold for the initial conditions and input sequences in the set $\Psi$ when considering sample-based \iIOSS/ on the input set $W$. This is stated in the following corollary.\\

\begin{cor}
	 Let system (\ref{eq:sys}) be \iIOSS/, let Assumption~\ref{ass:h} hold and consider some input set $W\subseteq (\mathbb{R}^q)^{\mathbb{I}_{\geq0}}$.   
	Moreover, consider  some set $K$ according to Definition~\ref{def:K}.
	Then, the system is sample-based  \iIOSS/ on $W$ with respect to $K$ if there exist functions $\gamma_w, \gamma_h \in \mathcal{K}$ and a finite time $t^*$ such that for any $(x_{01},x_{02},\boldsymbol{w_1},\boldsymbol{w_2})\in\Psi$, (\ref{eq:cond_sampling}) holds. \label{rem:condsuff}\\
\end{cor}
\par We now make the following assumption in order to show in Theorem~\ref{thm2} below that a necessary condition for sample-based \iIOSS/ is that (\ref{eq:cond_sampling}) holds for all$(x_{01},x_{02},\boldsymbol{w_1},\boldsymbol{w_2})\in\Psi$.\\

\begin{ass}
	System (\ref{eq:sys}) admits an \iIOSS/ bound as in (\ref{eq:iIOSS}) with $\beta \in  \mathcal{KL}$ for which  there exists  a time $T_{\beta}$ such that for each  $(x_{01},x_{02},\boldsymbol{w_1},\boldsymbol{w_2})\in \Psi$, there exists $t_{\psi}\leq T_{\beta}$ such that
	\begin{align}
		\begin{aligned}
			|\Delta x(t_{\psi})|
			&> \beta(| \Delta x_0
			|,t_{\psi}	).
		\end{aligned}
		\label{eq:Ass3}
	\end{align}
		\label{ass:Psixw}
\end{ass}
\par  Assumption \ref{ass:Psixw} means that for all  $(x_{01},x_{02},\boldsymbol{w_1},\boldsymbol{w_2})\in \Psi$ (i.e., pairs of initial conditions and inputs that do not result in a pair of i-ISS trajectories), the incremental asymptotic stability bound (compare (\ref{eq:Omega}) with the input-dependent term being omitted)
is violated before some (uniform) time $T_{\beta}$. 
To illustrate the meaning of Assumption \ref{ass:Psixw}, we use the following example for the case of no inputs affecting the system.
Consider a scalar, unstable linear system 
\begin{align}
	\begin{aligned}
		x(t+1)&=ax(t)\\
		y(t)&=x(t) 
	\end{aligned}
	\label{eq:exUnstable}
\end{align}
with $a>1$. The system is observable, and thus \iIOSS/ \cite{Knu20}. In fact, since we have full state measurements, we can choose any function $\beta \in \mathcal{KL}$ that satisfies $\beta(s,0)\geq s$ together with any $\gamma_2(s)\geq as$ and any $\gamma_1 \in \mathcal{K}$ for (\ref{eq:iIOSS}) to hold. Hence, one possible choice for  $\beta$ would be
\begin{align}
	\beta(|\Delta x_0|,t)=\begin{cases}
		c\sqrt{|\Delta x_0|}e^{-\lambda t},& |\Delta x_0|<1\\
		c|\Delta x_0|e^{-\lambda t},& |\Delta x_0|\geq 1
	\end{cases}	
	\label{eq:exampleBeta1}
\end{align}
with $c,\lambda\in \mathbb{R}_{>0},\  c\geq 1$. 
Due to $\beta$ being a square root function  in the first argument  (for $|\Delta x_0| < 1$), the time for which $|\Delta x(t)|$ exceeds $\beta(|\Delta x_0|,t)$  grows unboundedly as  \mbox{$|\Delta x_0|\rightarrow 0$}.  Hence, there exists no lower bound $T_{\beta}$, such that for all \footnote{Since system (\ref{eq:exUnstable}) has no inputs, we use the slight abuse of notation $(x_{01},x_{02}) \in \Psi$ instead of  $(x_{01},x_{02},\boldsymbol{w_1},\boldsymbol{w_2}) \in \Psi$ as in Definition \ref{def:Psi}.} $(x_{01},x_{02}) \in \Psi$, (\ref{eq:Ass3}) holds for some $t_{\psi} \leq T_{\beta}$. 
 Therefore, $\beta$ as in (\ref{eq:exampleBeta1}) does not make Assumption \ref{ass:Psixw} hold for all $(x_{01},x_{02})\in \Psi$. 
However,  as discussed above, also  any other $\mathcal{KL}$-function $\beta$  satisfying $\beta(s,0)\geq s$ can be used in (\ref{eq:iIOSS}), e.g., 
\begin{align}
	\beta(|\Delta x_0|,t)=c|\Delta x_0|e^{-\lambda t}, \quad c,\lambda \in \mathbb{R}_{\geq 0}, \  c\geq 1. 
	\label{eq:KLex}
\end{align}
Here, selecting $T_{\beta}>\frac{ln(c)}{ln(a)+\lambda}$ guarantees that for all $x_{01},x_{02} \in \mathbb{R}^n$, there exists some $t_{\psi}$ such that $|\Delta x(t)|>\beta(|\Delta x_0|,t)$. 
Therefore, with this choice of $\beta$, Assumption \ref{ass:Psixw} is satisfied. As can be seen by the given example, even if not every \iIOSS/ bound satisfies (\ref{eq:Ass3}), in many cases it is still possible to find an \iIOSS/ bound that does make (\ref{eq:Ass3}) hold. Thus, in the case of no inputs, Assumption \ref{ass:Psixw} is not overly restrictive in the sense that it does not exclude many system classes. 
\par
For the desired case of systems with inputs, we show below that in order to satisfy Assumption \ref{ass:Psixw}, the set of possible inputs in general needs to be constrained to some set $W \subseteq  (\mathbb{R}^q)^{\mathbb{I}_{\geq0}}$.
Consider again the unstable linear system from (\ref{eq:exUnstable}), but now with some added input, i.e., \mbox{$x(t+1)=ax(t)+w(t)$}. Choose $x_{01}=0,x_{02}\neq 0, w_1(t)=0, \forall t\geq 0$ and set $w_2(t)$ as  
\begin{align}
	w_2(t)=\begin{cases}
		(0.5-a)x(t),& 0\leq t<\bar{t},\\
		0,&t\geq\bar{t}.
	\end{cases}
\label{eq:stabinput}
\end{align}
Then, the input sequence $\boldsymbol{w_2}$ stabilizes the system for $t< \bar{t}$ and $|\Delta x(t)|$ exponentially converges to zero. However, this pair of trajectories is not a pair of \iISS/ trajectories because the system is unstable again after $t \geq \bar{t}$. Since $\bar{t}$ can be arbitrarily large, there exists no (uniform) time $T_{\beta}$ that makes (\ref{eq:Ass3}) hold for all $(x_{01}, x_{02},\boldsymbol{w_1}, \boldsymbol{w_2}) \notin \Lambda_{\iISS/}$.\\ 
\begin{rem}
	Notice that the choice of the functions $\beta, \gamma_1, \sigma_1$ influence the size of the set $\Psi$. Functions corresponding to a tighter \iIOSS/ bound yield a larger set $\Psi$ and vice versa.
	\par \textcolor{new}{Furthermore, one could also allow in Definition~\ref{def:Psi} for some joint and state-dependent set of input trajectories, i.e., $(\boldsymbol{w_1},\boldsymbol{w_1})\in W(x_{01},x_{02})\subseteq  (\mathbb{R}^q\times  \mathbb{R}^q)^{\mathbb{I}_{\geq0}}$, which may simplify the design of an input set  to satisfy Assumption~\ref{ass:Psixw}. However, for ease of presentation, we refrain from using this notation in the remainder of the paper.} \\
	\label{rem:Psi}
\end{rem}
In the following theorem, the aforementioned necessary condition for sample-based \iIOSS/  is presented.\\
\begin{thm}
	Let system (\ref{eq:sys}) be  sample-based \iIOSS/ with respect to some set $K$ according to Definition~\ref{def:K} and let Assumption \ref{ass:h} hold. Furthermore, consider an input set W such that Assumption 2 holds with $\beta = \bar{\beta}$, with $\bar{\beta}$ from (\ref{eq:s-iIOSS}).
Then, there exist  functions $\gamma_w, \gamma_h \in \mathcal{K}$  and a finite time instance $t^*$ 
	such that (\ref{eq:cond_sampling}) holds for all $(x_{01}, x_{02},\boldsymbol{w_1},\boldsymbol{w_2})  \in \Psi $. 
	\label{thm2}
\end{thm}
\begin{proof}
	Consider some functions $\bar{\beta} \in \mathcal{KL}$ and 
	$\bar{\gamma}_1,\bar{\gamma}_2 \in \mathcal{ K}$ satisfying the sample-based \iIOSS/ bound (\ref{eq:s-iIOSS}) and Assumption~\ref{ass:Psixw}. 
	For any $(x_{01}, x_{02},\boldsymbol{w_1},\boldsymbol{w_2}) \in \Psi$ it then holds from (\ref{eq:Ass3}) that  
	\begin{align}
		|\Delta x(t_{\psi})|
		> \bar{\beta}(| \Delta x_0
		|,t_{\psi}	)
	\end{align}
	for some \textcolor{rev}{$t_{\psi}\leq T_{\bar{\beta}}$}.
	Since the system is sample-based \iIOSS/ and, hence,
	\begin{align}
		\begin{aligned}
		|\Delta x(t_{\psi})|
		&\leq \bar{\beta}(| \Delta x_0
		|,t_{\psi}	)\oplus\bar{\gamma}_1(\sup_{0\leq \tau < t_{\psi}} | \Delta w(\tau)
		|) \\&\oplus \bar{\gamma}_2( \sup_{ \tau \in K_i, \tau < t_{\psi}} | \Delta h(\Delta x(\tau))|) \label{eq:sbiIOSStpsi}
		\end{aligned}
	\end{align}
	we know that
	\begin{align}
			\begin{aligned}
				\bar{\beta}(| \Delta x_0
				|,t_{\psi}	)&<\bar{\gamma}_1(\sup_{0\leq \tau < t_{\psi}} | \Delta w(\tau)
		|)\\&\oplus\bar{\gamma}_2( \sup_{ \tau \in K_i, \tau <t_{\psi}} | \Delta h(\Delta x(\tau))|). \label{eq:gamma2g}
			\end{aligned}
	\end{align}
	Using (i) the sample-based i-IOSS bound (\ref{eq:s-iIOSS}), (ii) inequality (\ref{eq:gamma2g}) and (iii) the fact that $\bar{\beta}$ is strictly decreasing in its second argument for a fixed $|\Delta x_0|$, it holds that
	\begin{align}
		\begin{aligned}
		|\Delta x(t)|&\leq \bar{\gamma}_1(\sup_{0\leq \tau < t} | \Delta w(\tau)
		|)\\&
		\oplus\bar{\gamma}_2(\sup_{ \tau \in K_i, \tau< t} | \Delta h(\Delta x(\tau))|), \quad \forall t\geq t_{\psi}.
		\end{aligned}
	\end{align}
	Recall that $t_{\psi}\leq T_{\bar{\beta}}$ for all $(x_{01},x_{02},\boldsymbol{w_1},\boldsymbol{w_2}) \in \Psi$. Thus, 
	\begin{align}\label{eq:thm2BoundX}
		\begin{aligned}
		&|\Delta x(t)|\leq \bar{\gamma}_1(\sup_{0\leq \tau < t} | \Delta w(\tau)
		|)\\
		&\oplus\bar{\gamma}_2(\sup_{ \tau \in K_i, \tau< t} | \Delta h(\Delta x(\tau))|), \ \forall t\geq T_{\bar{\beta}}, \ \forall K_i \in K.
		\end{aligned}
	\end{align}
	Since $h$ is $\mathcal{K}$-continuous, we can use (\ref{eq:Kcont}) and (\ref{eq:thm2BoundX}) to write
	\begin{align}
		|\Delta h(\Delta x(t))|&\leq \alpha_h(2\bar{\gamma}_1(\sup_{0\leq \tau < t} | \Delta w(\tau)
		|))\\&\oplus \alpha_h(2\bar{\gamma}_2(\sup_{ \tau \in K_i, \tau< t} |h(\Delta x (\tau))|)), \ \forall t\geq T_{\bar{\beta}}. \nonumber
	\end{align}
	Selecting $\gamma_w(\cdot)=\alpha_h(2\bar{\gamma}_1(\cdot))$, $\gamma_h(\cdot)=\alpha_h(2\bar{\gamma}_2(\cdot))$ and \makebox{$t^*= T_{\bar{\beta}}$} makes (\ref{eq:cond_sampling}) hold for all elements in $\Psi$.
\end{proof}
\par After having studied the notion of sample-based \iIOSS/ and having given conditions for an \iIOSS/ system to be sample-based \iIOSS/, we focus in the following section on extending our results to a formulation of \iIOSS/ that includes time-discounting terms. 

\section{Sample-based time-discounted \iIOSS/}
\label{sec:sbtdiISS}
The classical non-time-discounted \iIOSS/ formulation provides a bound characterized by the maximum of the norm of the input and output differences. In recent years, a time-discounted version of \iIOSS/ has been suggested that can advantageously be used to design nonlinear estimators. In particular, first suggestions  of discounting the influence of inputs and outputs  from the past on the bounds were made in \cite{All18,All19, Knu18} in the context of showing that a system admitting an input-ouput-to-state stability (IOSS) Lyapunov-function or  \iIOSS/ Lyapunov-function implies the system being IOSS or \iIOSS/, respectively \cite{All18,All19}  and for use in MHE \cite{Knu18}. Concerning the latter, using a time-discounted i-IOSS formulation and a corresponding time-discounted cost function in MHE as  in \cite{Knu23,Sch23} provides a bound on the estimation error that also  discounts the influence of the disturbances. Hence, robust stability directly implies that the estimation error converges to zero for vanishing disturbances. Additionally, the time-discounted cost function allows tighter upper bounds on the estimation error. Furthermore, it should be noted that the time-discounted \iIOSS/ formulation allows to transform a sum-based \iIOSS/ bound to a max-formulation as is discussed in \cite{Knu20}. Given the above motivation, an interesting question is to study a time-discounted \iIOSS/ notion also in a sample-based setting. This is particularly intuitive since it motivates the ongoing collection of meaningful measurements.
\par In this section, we first introduce a time-discounted version of sample-based \iIOSS/, then study conditions for an \iIOSS/ system to be sample-based time-discounted \iIOSS/ and address the relations between the two (discounted and non-discounted) notions of sample-based \iIOSS/.\\ 

\begin{defi}[Time-discounted \iIOSS/] 
	The system (\ref{eq:sys}) is time-discounted  \iIOSS/ if there exist functions $\beta_x,\beta_u,\beta_y \in \mathcal{KL}$  such that for any two initial
	conditions $x_{01}$, $x_{02}$ and any two input (disturbance) trajectories $\boldsymbol{w_1},\boldsymbol{w_2}$ the following holds for all $t \geq 0$
	\begin{align}
			|\Delta x(t)|
			&\leq \beta_x(| \Delta x_0
			|,t	)\oplus \max_{0\leq \tau< t}\beta_u(| \Delta w(\tau)
			|,t-\tau-1)\nonumber\\
			&\oplus\max_{0\leq \tau<t}\beta_y( |\Delta h(\Delta x(\tau))|,t-\tau-1).
			\label{eq:td-iIOSS}
	\end{align}
	\label{eq:td-\iIOSS/}
\end{defi}
\par As was recently shown, a system is time-discounted \iIOSS/ according to Definition~\ref{eq:td-\iIOSS/} if and only if it is \iIOSS/ \cite{All20}. In the following, we first formally define sample-based time-discounted \iIOSS/.  Afterwards, we show the following for the sample-based setting: (i) sample-based time-discounted i-IOSS implies sample-based i-IOSS; (ii) the sufficient condition (\ref{eq:cond_sampling}) for sample-based \iIOSS/ is also sufficient for sample-based time-discounted \iIOSS/; (iii) under Assumption \ref{ass:Psixw}, sample-based \iIOSS/ and sample-based time-discounted \iIOSS/ are equivalent (compare Fig. \ref{fig:schem}). Whether or not sample-based \iIOSS/ and sample-based time-discounted \iIOSS/ are equivalent also without Assumption \ref{ass:Psixw} is an interesting open issue for future research.
\begin{figure}[!t]
	\centering
	\setlength\figurewidth{0.8\columnwidth} 
	\centerline{\input{figures/overview}}
	\caption{Schematic sketch of the main results, visualizing implications between \iIOSS/, sample-based  \iIOSS/ and  sample-based time-discounted  \iIOSS/. Conditions or Assumptions for which the respective statement  holds true are given in square brackets.} 
	\label{fig:schem}
\end{figure}
\par Before showing the aforementioned, we first define a sample-based version of time-discounted \iIOSS/.\\
\begin{defi}[Sample-based time-discounted \iIOSS/]
	Consider some set $K$ according to Definition~\ref{def:K}. The system~(\ref{eq:sys}) is sample-based time-discounted \iIOSS/ with respect to $K$ if there exist functions $\bar{\beta}_x,\bar{\beta}_u,\bar{\beta}_y \in \mathcal{KL}$  such that for any two initial
	conditions $x_{01}$, $x_{02}$ and any two input (disturbance) trajectories $\boldsymbol{w_1},\boldsymbol{w_2}$ the following holds for any $K_i  \in K$ and all $t \geq 0$ 
	\begin{align}
			|
			\Delta x(t)
			|
			&\leq \bar{\beta}_x(| \Delta x_0
			|,t	)\oplus\max_{0\leq \tau< t}\bar{\beta}_u(| \Delta w(\tau)
			|,t-\tau-1) \nonumber\\
			&\oplus \max_{ \tau \in K_i, \tau< t}\bar{\beta}_y( |\Delta h(\Delta x(\tau))|,t-\tau-1).
			\label{eq:s-td-iIOSS}
	\end{align}
\label{def:sbtdiIOSS}
\end{defi}
The following relationship between Definition~\ref{def:sbiIOSS} and Definition~\ref{def:sbtdiIOSS} is straightforwardly obtained.\\
\begin{lem}
	If a system is sample-based time-discounted \iIOSS/, then it is sample-based \iIOSS/.
	\label{lem}
\end{lem}
\begin{proof}
	Suppose the system is sample-based time-discounted \iIOSS/ with functions $ \bar{\beta}_x,\bar{\beta}_u,\bar{\beta}_y \in \mathcal{KL}$ satisfying the sample-based time-discounted \iIOSS/ bound. Then, selecting
	$\bar{\beta}(s,r)=\bar{\beta}_x(s,r)$, $\bar{\gamma}_1(s)=\bar{\beta}_u(s,0)$ and $\bar{\gamma}_2(s)=\bar{\beta}_y(s,0)$ makes (\ref{eq:s-iIOSS}) hold.
\end{proof}
\par The sufficient condition in Theorem~\ref{thm1}  for system (\ref{eq:sys}) to be sample-based \iIOSS/  is also a sufficient condition for sample-based time-discounted \iIOSS/. 
Before stating the corresponding theorem, we introduce first the following lemma that will be used to prove the aforementioned.\\
\begin{lem}
Consider a set $K$ according to Definition \ref{def:K}. If $t^{i+j}_k \in K_{i+j}$ for some $i,j,k \in \mathbb{I}_{>0}$, then $t^{i+j}_k + t^i_j \in K_i$.
\label{lem:K}
\end{lem}
\begin{proof}
The proof follows from Definition \ref{def:K}. We can express $t^{i+j}_k = \sum_{r=i+j}^{i+j+k-1} \delta_r$. Now it is clear that 
\begin{align}
	\begin{aligned}
t^{i+j}_k + t^i_j &= \sum_{r=i+j}^{i+j+k-1} \delta_r + \sum_{r=i}^{i+j-1} \delta_r\\ &= \sum_{r=i}^{i+j+k-1} \delta_r\\ &= t^i_{i+j+k} \in K_i.
\end{aligned}
\end{align}
\end{proof}
\begin{thm}
	Let system (\ref{eq:sys}) be \iIOSS/  and let Assumption~\ref{ass:h} hold.  
	Moreover, consider  some set $K$ according to Definition~\ref{def:K}.
	Then, the system is sample-based time-discounted \iIOSS/ with respect to $K$ if there exist functions $\gamma_w, \gamma_h \in \mathcal{K}$ and a finite time $t^*$ such that for any two initial conditions $x_{01}$, $x_{02}$ and any two input (disturbance) trajectories $\boldsymbol{w_1},\boldsymbol{w_2}$, (\ref{eq:cond_sampling}) holds. 
	\label{thm1_td}
\end{thm}
\begin{proof} 
	The proof is divided into two parts. Part one focuses on designing functions satisfying (\ref{eq:s-td-iIOSS}) for $t<t^*$ and part two addresses the case of $t\geq t^*$. Note that we can choose the value of $t^*$ large enough such that every $t^*$-long interval contains at least one measurement instance, since the measurements are separated by a maximum of $\delta_{max}$ time units (compare Definition~\ref{def:K}). We will exploit this in the second part of the proof.
	\par {\it Part one:}
	Theorem~\ref{thm1} states that if (\ref{eq:cond_sampling}) holds then the system is sample-based \iIOSS/, i.e.,
	\begin{align}
		\begin{aligned}
		&	|
			\Delta x(t)
			|
			\leq \bar{\beta}(| \Delta x_0
			|,t	)\oplus \bar{\gamma}_1(\sup_{0\leq \tau < t}  | \Delta w(\tau)
			|)\\
			&\oplus \bar{\gamma}_2(\sup_{ \tau \in K_i, \tau <t} |\Delta h(\Delta x(\tau))|), \forall t\geq 0, \ \forall K_i \in K.
		\end{aligned}
	\end{align}
Note that since the system is \iIOSS/, it is also time-discounted \iIOSS/ \cite{All20}, i.e., (\ref{eq:td-iIOSS}) holds.
	We argue that there exist functions $\tilde{\beta}_x, \tilde{\beta}_u$ and $\tilde{\beta}_y \in \mathcal{KL}$ such that for $t<t^*$ it holds that 
	\begin{align}
		|
		\Delta x(t)
		|
		&\leq \tilde{\beta}_x(| \Delta x_0
		|,t	)\oplus\ \max_{ 0\leq \tau < t} \tilde{\beta}_u( | \Delta w(\tau)
		|, t-\tau-1)\nonumber \\ &\oplus \max_{\tau \in K_i, \tau<t}\tilde{\beta}_y( | \Delta h(\Delta x(\tau))
		|, t-\tau-1) , \label{eq:betatildexu}\\ 
		&\phantom{\qquad \qquad \qquad \quad \ } \forall t \in [0,t^*), \ \forall K_i\in K .\nonumber
	\end{align}
\textcolor{rev}{	If $t^*<2$, then (\ref{eq:betatildexu}) holds for $t<t^*$  for 	$\tilde{\beta}_x(s,r)=\bar{\beta}(s,r)$ and any choice of 	$\tilde{\beta}_u(s,r)$ and 	$\tilde{\beta}_y(s,r)$.
	Therefore, we consider in the following only the case of $t^*\geq 2$.
	The existence of functions $\tilde{\beta}_x, \tilde{\beta}_u$ and $\tilde{\beta}_y \in \mathcal{KL}$ such that (\ref{eq:betatildexu}) holds can be demonstrated by selecting the functions in question, for instance, in the following manner}
	\begin{subequations}\label{eq:betatilde}
		\begin{alignat}{3}
			\tilde{\beta}_x(s,r)&=\bar{\beta}(s,r)  \label{eq:betatildex}\\ 
			\tilde{\beta}_u(s,r)&=\frac{1}{\sigma_u(t^*-2)}\bar{\gamma}_1(s)\sigma_u(r) \label{eq:betatildeu}\\
			\tilde{\beta}_y(s,r)&=\frac{1}{\sigma_y(t^*-2)}\bar{\gamma}_2(s)\sigma_y(r) \label{eq:betatildey}
		\end{alignat}
	\end{subequations}
with $\sigma_u, \sigma_y$ being any $\mathcal{L}$-functions. Since $\mathcal{L}$-functions are strictly decreasing, the minima of \mbox{$\sigma_u(t-\tau-1), \sigma_y(t-\tau-1)$} for $t\in [0,t^*)$ and $\tau \in [0,t)$ are reached for $\tau=0$ and $t=t^*-1$, i.e., $\sigma_u(t^*-2)$ and $\sigma_y(t^*-2)$. 
Hence, by  (\ref{eq:betatildeu}) and (\ref{eq:betatildey}) it holds for all $t<t^*$ that \begin{subequations}
		\begin{alignat}{2}
			\max_{ 0\leq \tau < t} &\tilde{\beta}_u( | \Delta w(\tau)|,t-\tau-1)\geq\bar{\gamma}_1(\sup_{0\leq \tau < t} |  \Delta w(\tau)|),\\
			\begin{split}
			 \max_{\substack{ \tau \in K_i,\\ \tau< t}}&\tilde{\beta}_y( | \Delta h(\Delta x(\tau))|,t-\tau-1)\\\geq&\bar{\gamma}_2(\sup_{\substack{ \tau \in K_i,\\ \tau< t}} |\Delta h(\Delta x(\tau))|).
			 \end{split}
			 	\end{alignat}
	 \end{subequations}
	
	\par {\it Part two:}
	By applying (\ref{eq:cond_sampling}), $|\Delta h(\Delta x(t))|$ can be bounded as follows for $t\geq t^*$
	\begin{align}
		|\Delta h(\Delta x(t))|&\leq \max_{0\leq \tau < t}\beta_w ( | \Delta w(\tau)
		|,t-\tau-1) \label{eq:thm3_h}\\
		&\oplus \max_{ \substack{ \tau \in K_i,\\ \tau < t}}\beta_h(  | \Delta h(\Delta x(\tau))|,t-\tau-1), \  \forall   t\geq t^* \nonumber
	\end{align}
	with \begin{subequations}
		\begin{alignat}{2}
			\beta_w(s,r)&=\frac{1}{\sigma(2t^*-1)}\gamma_w(s)\sigma(r),	\label{eq:thm3Betaw}\\ 
			\beta_h(s,r)&=\frac{1}{\sigma(2t^*-1)}\gamma_h(s)\sigma(r), \quad \sigma \in \mathcal{L}.\label{eq:thm3Betah}
		\end{alignat}
		\label{eq:thm3Beta}
	\end{subequations}
That (\ref{eq:thm3Beta}) together with (\ref{eq:cond_sampling})  make (\ref{eq:thm3_h}) hold for $t^*\leq t\leq 2 t^*$ can be easily seen by the same arguments as in part one of the proof.  In the following, we show by contradiction that (\ref{eq:thm3_h}) also holds in case of $t> 2t^*$. Assume there exist trajectories $x(t;x_{01},\boldsymbol{w_1}), x(t;x_{02},\boldsymbol{w_2})$ and a time instance $\bar{t}> 2t^*$ such that (\ref{eq:thm3_h}) does not hold, i.e.,
\begin{align}
\begin{aligned}
	|\Delta h(\Delta x(\bar{t}))|&> \max_{0\leq \tau < \bar{t}}\beta_w ( | \Delta w(\tau)
	|,\bar{t}-\tau-1) \\
	&\oplus \max_{ \tau \in K_p, \tau < \bar{t}}\beta_h(  | \Delta h(\Delta x(\tau))|,\bar{t}-\tau-1)
\end{aligned}
\label{eq:ass_cont}
\end{align}
for some $K_p \in K$.
Next, consider a pair of initial states and input trajectories  ($\tilde{x}_{01},\tilde{x}_{02}, \boldsymbol{\tilde{w}_1},\boldsymbol{\tilde{w}_2}$) with \begin{align}
	\begin{aligned}
	\tilde{x}_{01}&= x(\bar{t}-\zeta;x_{01},\boldsymbol{w_1}),\\
	 \tilde{x}_{02}&= x(\bar{t}-\zeta;x_{02},\boldsymbol{w_2}) 
	\end{aligned}
\label{eq:newIni}
\end{align}
 and with $\zeta$ being some positive integer less than $\bar{t}$. Furthermore, $\tilde{\boldsymbol{w}}$ is given by taking the sequence $\boldsymbol{w}$ starting from $\bar{t}-\zeta$, i.e., $\tilde{w}(t)=w(\bar{t}-\zeta+t)$ for all $t\geq0$. Hence, 
\begin{align}
	\begin{aligned}
		&x(\bar{t}-\zeta+t;x_{01},\boldsymbol{w}_1)-x(\bar{t}-\zeta+t;x_{02},\boldsymbol{w}_2)\\=& x(t;\tilde{x}_{01},\boldsymbol{\tilde{w}}_1)-x(t;\tilde{x}_{02},\boldsymbol{\tilde{w}}_2),  \quad \forall t\geq0
	\end{aligned}
\end{align} 
abbreviated by
 \begin{align}
 	\Delta x(\bar{t}-\zeta+t)=\Delta \tilde{x}(t), \quad \forall t\geq0. \label{eq:shiftTraj}
 \end{align}
Since our choice of $t^*$ ensures that there exist at least one sampling instance in the interval $[pt^*, (p+1)t^*)$ for all $p\in \mathbb{I}_{\geq0}$, we can select a $\zeta$ for which it holds that $\bar{t}-\zeta \in K_p$ and $t^*\leq\zeta<2t^*$. This means that we can write $K_p = \{t_1^p, t_2^p,..., t_{j-1}^p, t_j^p,\ldots\}$, where $t_j^p = \bar{t} - \zeta$ for some $j$.
Due to $\zeta \in [t^*, 2t^*)$, 
as discussed above (compare (\ref{eq:thm3_h})), it holds that 
\begin{align}
	\begin{aligned}
	|\Delta h(\Delta\tilde{x}(\zeta))|\leq \max_{0\leq \tau < \zeta}\beta_w ( | \Delta \tilde{w}(\tau)
	|,\zeta-\tau-1)\\
	\oplus \max_{\substack{ \tau \in K_{i},\\ \tau < \zeta}}\beta_h(  | \Delta h(\Delta \tilde{x}(\tau))|,\zeta-\tau-1), \quad \forall K_i\in K
	\end{aligned}
\label{eq:tilde_h_bound}
\end{align}
with $\beta_w$ and $\beta_h$ as in (\ref{eq:thm3Betaw}) and (\ref{eq:thm3Betah}). 
Notice that (\ref{eq:tilde_h_bound}) holds for all $K_i \in K$. Here, it will be convenient to consider $K_i = K_{p+j}$, with $j$ defined from the above discussed equality $t_j^p = \bar{t} - \zeta \in K_p$.
Denote by $\tau_{\text{max-w}}$ the value of $\tau$ that maximizes the first term in the right-hand side of (\ref{eq:tilde_h_bound}). Similarly, let $\tau_{\text{max-h}}$ denote the value of $\tau$ that maximizes the second term in the right-hand side of (\ref{eq:tilde_h_bound}) for the case $K_i = K_{p+j}$
Due to (\ref{eq:newIni}) and (\ref{eq:shiftTraj}) we know that 
\begin{align}
	\begin{aligned}
		&\beta_w(|\Delta \tilde{w}(\tau_{\text{max-w}})
		|,\zeta-\tau_{\text{max-w}}-1)\\
	= 	&\beta_w(|\Delta w(\tau_{\text{max-w}}+\bar{t}-\zeta))
	|,\zeta-\tau_{\text{max-w}}-1) 
	\end{aligned}
\end{align}
and 
\begin{align}
	\begin{aligned}
		&\beta_h(|\Delta h(\Delta \tilde{x}(\tau_{\text{max-h}}))
		|,\zeta-\tau_{\text{max-h}}-1)\\
		=&\beta_h(|\Delta h(\Delta x(\tau_{\text{max-h}}+\bar{t}-\zeta))
		|,\zeta-\tau_{\text{max-h}}-1) .
	\end{aligned}
\end{align}
Note that $\zeta-\tau_{\text{max-w}}-1=\bar{t}-(\tau_{\text{max-w}}+\bar{t}-\zeta)-1$ and analogously the same holds for $\tau_{\text{max-h}}$.
Hence, from (\ref{eq:tilde_h_bound})
\begin{align}
		&|\Delta h(\Delta x(\bar{t}))| \label{eq:thm3contra} \\&\leq \beta_w(|\Delta w(\tau_{\text{max-w}}+\bar{t}-\zeta)
		|,\bar{t}-(\tau_{\text{max-w}}+\bar{t}-\zeta)-1) \nonumber\\
		&\oplus\beta_h(|\Delta h(\Delta x(\tau_{\text{max-h}}+\bar{t}-\zeta))
		|,\bar{t}-(\tau_{\text{max-h}}+\bar{t}-\zeta)-1).\nonumber 
\end{align}
In order to finish our contradiction argument, we must show that $\tau_{max-h} + \bar{t} - \zeta$ is an element of $K_p$. This is indeed shown by Lemma \ref{lem:K}, by noticing that $\tau_{max-h} \in K_{p+j}$ and $t^p_j = \bar t - \zeta \in K_p$. 
Therefore, (\ref{eq:thm3contra}) contradicts (\ref{eq:ass_cont}).
\par	After proving (\ref{eq:thm3_h}), we can now apply $\beta_y(\cdot,0)$ to both sides of (\ref{eq:thm3_h}) to obtain 
	\begin{align}
		\begin{aligned}
			&	\beta_y( |\Delta h(\Delta x(t))|,0)\\
			&\leq \max_{ \tau \in K_i, \tau < t}\beta_y(\beta_h(  | \Delta h(\Delta x(\tau))|,t-\tau-1),0)\\ &\oplus\max_{0\leq \tau < t}\beta_y(\beta_w ( | \Delta w(\tau)
			|,t-\tau-1),0), \\ &\forall t\geq t^*, \ \forall K_i \in K.  	\label{eq:th3boundmaxy1}
		\end{aligned}
	\end{align} 
	Since (\ref{eq:th3boundmaxy1}) holds for all $t\geq t^*$ and $\beta_y$ is non-increasing in its second argument, we can write the following upper bound for $\max_{t^*\leq \tau<t}\beta_y( |\Delta h(\Delta x(\tau))|,t-\tau-1)$ for $t\geq t^*$ and any $K_i \in K$
	\begin{align}
		\begin{aligned}
			&	\max_{t^*\leq \tau<t}\beta_y( |\Delta h(\Delta x(\tau))|,t-\tau-1)\\&\leq \max_{t^*\leq \tau<t}\beta_y( |\Delta h(\Delta x(\tau))|,0)\\
			&\stackrel{(\ref{eq:th3boundmaxy1})}{\leq} \max_{ \tau \in K_i, \tau < t}\beta_y(\beta_h(  | \Delta h(\Delta x(\tau))|,t-\tau-1),0)\\ &\oplus\max_{0\leq \tau < t}\beta_y(\beta_w ( | \Delta w(\tau)
			|,t-\tau-1),0) .  	\label{eq:th3boundmaxy}
		\end{aligned}
	\end{align} 
	 Notice that we can split the output-dependent term of the time-discounted \iIOSS/ bound in (\ref{eq:td-iIOSS}) as follows
	 \begin{align}
	 	\begin{aligned}
	 		|
	 	\Delta x(t)
	 	|
	 	&\leq \bar{\beta}_x(| \Delta x_0
	 	|,t	)\oplus\max_{0\leq \tau< t}\bar{\beta}_u(| \Delta w(\tau)
	 	|,t-\tau-1)\\
	 	&\oplus \max_{ 0\leq\tau< t^*}\bar{\beta}_y( |\Delta h(\Delta x(\tau))|,t-\tau-1)\\ &\oplus \max_{  t^*\leq \tau<t}\bar{\beta}_y( |\Delta h(\Delta x(\tau))|,t-\tau-1), \\ & \forall t\geq 0, \ \forall K_i\in K.
	 	\end{aligned}
	 	\label{eq:tdsplit}
	 \end{align}
	  An upper bound for \mbox{$\max_{t^*\leq \tau<t}\beta_y( |\Delta h(\Delta x(\tau))|,t-\tau-1)$} is already shown in  (\ref{eq:th3boundmaxy}). Now it remains to  upper bound $\max_{ 0\leq\tau< t^*}\bar{\beta}_y( |\Delta h(\Delta x(\tau))|,t-\tau-1)$.
	  Applying the results from part one (i.e., inequality (\ref{eq:betatildexu}))  and $\mathcal{K}$-continuity of $h$, we obtain the following bound  for all $t<t^*$ and any $K_i\in K$
	  \begin{align}
	  	|\Delta h(\Delta x(t))|&\leq \alpha_h( \tilde{\beta}_x(| \Delta x_0
	  	|,t	)) \nonumber\\&\oplus\ \alpha_h(\max_{ 0\leq \tau < t} \tilde{\beta}_u( | \Delta w(\tau)
	  	|, t-\tau-1)) \\ &\oplus \alpha_h(\max_{\tau \in K_i, \tau<t}\tilde{\beta}_y( | \Delta h(\Delta x(\tau))
	  	|, t-\tau-1)).\nonumber
	  \end{align}
	Using the same arguments as for deriving (\ref{eq:th3boundmaxy1}) and (\ref{eq:th3boundmaxy}) yields
	 	\begin{align}
	 		\begin{aligned}
	 			&	\max_{ 0\leq\tau< t^*}\beta_y( |\Delta h(\Delta x(\tau))|,t-\tau-1)\\
	 			&\leq \beta_y(\alpha_h( \tilde{\beta}_x(| \Delta x_0
	 			|,t	),0) \\&\oplus\ \max_{ 0\leq \tau < t}\beta_y(\alpha_h( \tilde{\beta}_u( | \Delta w(\tau)
	 			|, t-\tau-1)),0)\\ &\oplus \max_{\tau \in K_i, \tau<t}\beta_y(\alpha_h(\tilde{\beta}_y( | \Delta h(\Delta x(\tau))
	 			|, t-\tau-1)),0).	\label{eq:bft*}
	 		\end{aligned}
	 	\end{align} 	 
By inserting (\ref{eq:th3boundmaxy}) and (\ref{eq:bft*}) in (\ref{eq:tdsplit}) we obtain an  upper bound for $|\Delta x(t)|$ for $t\geq t^*$ and any $K_i\in K$.
Combining the results form part one and two of the proof, we can select the following functions to make (\ref{eq:s-td-iIOSS}) hold for all $t\geq 0$
\begin{subequations}
		\begin{alignat}{3}
			\bar{\beta}_x(s,r)& = \tilde{\beta}_x(s,r) \oplus \beta_y(\alpha_h( \tilde{\beta}_x(s,r),0)\\
			\begin{split}
			\bar{\beta}_u(s,r)& =  \beta_y(\beta_w(s,r),0)\oplus\tilde{\beta}_u(s,r)\\&\oplus \beta_y(\alpha_h( \tilde{\beta}_u(s,r),0))\end{split}\\
			\begin{split}
			\bar{\beta}_y(s,r)&=\beta_y(\beta_h(s,r),0)\oplus\tilde{\beta}_y(s,r)\\&\oplus \beta_y(\alpha_h(\tilde{\beta}_y(s,r),0)).
			\end{split}
				\end{alignat}
	\end{subequations}
\end{proof}
	
Next we show that if Assumption \ref{ass:Psixw} holds for some set $W$, then sample-based \iIOSS/ and sample-based time-discounted \iIOSS/ imply each other on the input set $W$. This means that the necessary condition for sample-based \iIOSS/ (cf. Theorem~\ref{thm2}) is also a necessary condition for sample-based time-discounted \iIOSS/. To obtain this result we use Sontag's $\mathcal{KL}$-function lemma.\\
\begin{lem}[Sontag's $\mathcal{KL}$-function lemma \cite{Son98}]
	For  every $\beta \in  \mathcal{KL}$ there  exist $\alpha_1,\alpha_2\in \mathcal{K}_{\infty}$ such that 
	\begin{align}
		\alpha_1(\beta(s,r))\leq \alpha_2(s)e^{-r} 
	\end{align}
	for all $s\in \mathbb{R}_{\geq0}$ and $r\in \mathbb{I}_{\geq 0}$.\\
	\label{lem:KL}
\end{lem}
\par Furthermore, to establish the before mentioned relation between sample-based \iIOSS/ and sample-based time-discounted \iIOSS/ 
\textcolor{rev}{we must assume shift-invariance of the input set $W$, as it is stated in the following assumption.}\\

 \begin{ass}
 Consider an input sequence $\boldsymbol{w}=\{w(0),w(1),\ldots\}$. If $\boldsymbol{w}\in W$, then $\sigma^i \boldsymbol{w} \in W$ with $\sigma^i \boldsymbol{w}$ denoting the shifted sequence $\{w(i), w(i+1), \ldots\}$.\\
 \label{ass:shiftinv}
 \end{ass}
\begin{thm}
Consider system (\ref{eq:sys}) and let Assumption \ref{ass:h} hold. Furthermore, consider an input set $W$ such that Assumption~2 holds with  $\beta =\bar{\beta}$  with $\bar{\beta}$ from (\ref{eq:s-iIOSS}).
	 Moreover, $W$ satisfies Assumption \ref{ass:shiftinv}. 
	Then, system (\ref{eq:sys}) is  sample-based time-discounted \iIOSS/ on $W$ if and only if it is sample-based \iIOSS/ on $W$.
\end{thm}
\begin{proof}
	By Lemma \ref{lem} we know already that sample-based time-discounted \iIOSS/ implies sample-based \iIOSS/. It remains to be shown that under Assumption \ref{ass:Psixw}, sample-based \iIOSS/ also implies sample-based time-discounted \iIOSS/ on the input set $W$. By Definition \ref{def:Psi} we know that for all $(x_{01},x_{02},\boldsymbol{w_1},\boldsymbol{w_2})\notin \Psi$ with $\boldsymbol{w_1},\boldsymbol{w_2}\in W$,  there exists a time-discounted bound  of the form
	\begin{align}
	\begin{aligned}
	|\Delta x(t)|
	&\leq \bar{\beta}(| \Delta x_0
	|,t	)\\&\oplus \max_{0\leq \tau< t}\bar{\gamma}_1(| \Delta w(\tau)
	|) \sigma_1(t-\tau-1), \quad \forall t\geq 0. \label{eq:tdiISS}
\end{aligned}
	\end{align}
\par 	
Now we determine the bounds of $|\Delta x(t)|$  for the elements of the set $\Psi$. For this, we will divide the rest of the proof into two parts. Part one addresses the case of $t\leq 2T_{\bar{\beta}}$ with $T_{\bar{\beta}}$ as in Assumption \ref{ass:Psixw}. The second part focuses on the case of 
 $t> 2 T_{\bar{\beta}}$, before finally combining all previous results to derive an upper  bound of $|\Delta x(t)|$ for all $t\geq 0$. 
 \par {\it Part one:}
 Since the system is sample-based \iIOSS/ on $W$, from Assumption \ref{ass:Psixw} it follows that for all \mbox{$(x_{01},x_{02},\boldsymbol{w_1},\boldsymbol{w_2})\in \Psi$}
 \begin{align}
 	\begin{aligned}
 		|\Delta x(t)|&\leq \bar{\gamma}_1(\sup_{0\leq \tau < t} | \Delta w(\tau)
 		|)\\
 		&\oplus\bar{\gamma}_2(\sup_{ \tau \in K_i, \tau< t} | \Delta h(\Delta x(\tau))|), \\ &\forall t\geq T_{\bar{\beta}}, \ \forall K_i \in K.
 	\end{aligned}
 	\label{eq:11sb}
 \end{align}
 In the following we want to show that there also exist functions $\bar{\beta}_u, \bar{\beta}_y \in \mathcal{KL}$ such that for all $(x_{01},x_{02},\boldsymbol{w_1},\boldsymbol{w_2})\in \Psi$ 
 \begin{align}
 	\begin{aligned}
 		|\Delta x(t)|	&\leq \max_{0\leq \tau < t}\bar{\beta}_u ( |\Delta w(\tau)|,t-\tau-1)\\ &\oplus \max_{\substack{ \tau \in K_i,\\ \tau < t}}\bar{\beta}_y(  | \Delta h(\Delta x(\tau))|,t-\tau-1), \\ &\forall t\geq T_{\bar{\beta}}, \ \forall K_i \in K.
 		\label{eq:beta12_mod}
 	\end{aligned}
 \end{align}
 It will be convenient to first define the following functions \mbox{$\beta_1$ and $\beta_2$}
 \begin{subequations}
 	\label{eq:betas}
 	\begin{align}
 		\begin{split}
 			\beta_1&(|\Delta w(\tau)| ,t-\tau-1)\\&=\bar{\gamma}_1(|\Delta w(\tau)|) \sigma(t-\tau-1)\frac{1}{\sigma(2T_{\bar{\beta}}-1 )}, \label{eq:beta1}
 		\end{split}\\
 	\begin{split}
 		\beta_2&(|\Delta h(x(\tau))| ,t-\tau-1)\\&=\bar{\gamma}_2(|\Delta h(x(\tau))|) \sigma(t-\tau-1)\frac{1}{\sigma(2T_{\bar{\beta}}-1 )},
 		\label{eq:beta2}
 	\end{split}
 \end{align}
 \end{subequations}
 for an arbitrary $\sigma \in \mathcal{L}$.
 Selecting  $\bar{\beta}_u =\beta_1$ and $\bar{\beta}_y=\beta_2$ 
 makes (\ref{eq:beta12_mod}) hold for the case of $t \in  [T_{\bar{\beta}},2T_{\bar{\beta}}]$. This can be easily seen by using the same arguments as in part one of the proof of  Theorem~\ref{thm1_td}. 
The following combination of $\beta_1, \beta_2$ and $\bar{\beta}$ upper bounds $|\Delta x(t)|$ for all $(x_{01},x_{02},\boldsymbol{w_1},\boldsymbol{w_2})\in \Psi$ for the case of  $t\leq 2T_{\bar{\beta}}$ 
	\begin{align}
		\begin{aligned}
		|\Delta x(t)|	\leq \bar{\beta}(|\Delta x_0|,t)\oplus \max_{0\leq \tau < t}\beta_1 ( |\Delta w(\tau)|,t-\tau-1)\\ \oplus \max_{\substack{ \tau \in K_i,\\ \tau < t}}\beta_2(  | \Delta h(\Delta x(\tau))|,t-\tau-1),  \ \forall K_i \in K.
		\label{eq:PsixwlessT}
	\end{aligned}
	\end{align} 
 \par {\it Part two:}
	Now it remains to be shown that there exist functions that make (\ref{eq:beta12_mod}) hold for all $(x_{01},x_{02},\boldsymbol{w_1},\boldsymbol{w_2})\in \Psi$  for the case of $t >  2T_{\bar{\beta}}$. 
	For this, we first need to distinguish between pairs of trajectories generated by the elements in  $\Psi$ that for $t \rightarrow \infty$ (i) converge to each other or are bounded by their disturbances $|\Delta \boldsymbol{w}|$ and (ii) those for which both is not the case. 
	Suppose $(x_{01},x_{02},\boldsymbol{w_1},\boldsymbol{w_2})$  is in $\Psi$.  \mbox{Moreover},  consider the solution at some time  $t_0 \in \mathbb{I}_{> 0}$ as a further pair of initial states $(x(t_0;x_{01},\boldsymbol{w_1}), x(t_0;x_{02},\boldsymbol{w_2}))$  denoted by $\tilde{x}_{01},\tilde{x}_{02}$, such that 
	\begin{align}
		\begin{aligned}
			x(t_0+t;x_{01},\boldsymbol{w}_1)&=x(t;\tilde{x}_{01},\boldsymbol{\tilde{w}}_1),\\
			x(t_0+t;x_{02},\boldsymbol{w}_2)&=x(t;\tilde{x}_{02},\boldsymbol{\tilde{w}}_2), \quad\forall t\geq0,
		\end{aligned}
	\end{align} 
where $\tilde{w}_i(t)=w_i(t_0+t)$ for $i=1,2$. Recall that, by Assumption \ref{ass:shiftinv}, $\tilde{\boldsymbol{w}}_i \in W$.
	If for all $t_0$ it holds that $(x(t_0;x_{01},\boldsymbol{w_1}), x(t_0;x_{02},\boldsymbol{w_2}),\boldsymbol{\tilde{w}}_1,\boldsymbol{\tilde{w}}_2) \in \Psi$ then  $(x_{01},x_{02},\boldsymbol{w_1},\boldsymbol{w_2})$ defines a pair of trajectories in case (ii). Consequently, it can be seen that (\ref{eq:betas})  makes  (\ref{eq:beta12_mod}) also hold for $t>2 T_{\bar{\beta}}$ by using the same arguments as for proving (\ref{eq:thm3_h}) for  $t>2t^*$ in part two of the proof of Theorem \ref{thm1_td}.  
	In case (i), we define the time instance $\hat{t}$ such that  $t_0 = \hat{t}$ is the first time instance for which $(x(t_0;x_{01},\boldsymbol{w_1}), x(t_0;x_{02},\boldsymbol{w_2}),\boldsymbol{\tilde{w}}_1,\boldsymbol{\tilde{w}}_2) \notin \Psi$, i.e., $(x(t_0;x_{01},\boldsymbol{w_1}), x(t_0;x_{02},\boldsymbol{w_2}),\boldsymbol{\tilde{w}}_1,\boldsymbol{\tilde{w}}_2) \in \Psi$ for all $t_0< \hat{t}$. This means that, from $\hat{t}$ on, the two solutions $x(t;x_{01},\boldsymbol{w_1}),x(t;x_{02},\boldsymbol{w_2})$ either converge to each other or differ only due to their input difference $|\Delta \boldsymbol{w}|$. Therefore, for $2T_{\bar{\beta}}<t< \hat{t}+2 T_{\bar{\beta}}$  it can still be shown that (\ref{eq:betas})  makes  (\ref{eq:beta12_mod}) hold by again using the same arguments as in the beginning of part  two of the proof of Theorem~\ref{thm1_td}. 
 \par In the following we want to show   that there exist functions that make (\ref{eq:beta12_mod}) hold for the case of $t >  \hat{t}+2 T_{\bar{\beta}}$ for all $(x_{01},x_{02},\boldsymbol{w_1},\boldsymbol{w_2})\in \Psi$ that behave according to case (i). For this, note from (\ref{eq:tdiISS}) that  $|\Delta x(t)|$ is bounded as follows for $t\geq \hat{t}$
	\begin{align}
	\begin{aligned}
		&|\Delta x(t)|\leq \bar{\beta}(|\Delta x(\hat{t})|,t-\hat{t}) \\ &\oplus \max_{\hat{t}\leq \tau < t}\bar{\gamma}_1(| \Delta w(\tau)
		|) \sigma_1(t-\tau-1), \ \forall t\geq \hat{t}.
	\end{aligned}
	\label{eq:DeltaX_c2}
\end{align}
Applying Lemma \ref{lem:KL} to $\bar{\beta}$ we obtain
\begin{align}
	\alpha_1(\bar{\beta}(|\Delta x(\hat{t})|,t-\hat{t}))\leq \alpha_2(|\Delta x(\hat{t})|)e^{-(t-\hat{t})}
	\label{eq:SontagApp}
\end{align}
for some $\alpha_1,\alpha_2 \in \mathcal{K}$.
Therefore, applying $\alpha_1$ to both sides of (\ref{eq:DeltaX_c2}) yields
\begin{align}
	\begin{aligned}
		&\alpha_1(|\Delta x(t)|)\leq \alpha_2(|\Delta x(\hat{t})|)e^{-(t-\hat{t})} \\&\oplus \max_{\hat{t}\leq \tau < t}\alpha_1(\bar{\gamma}_1(| \Delta w(\tau)
		|) \sigma_1(t-\tau-1)), \ \forall t\geq \hat{t}.
	\end{aligned}
\end{align}
Since $|\Delta x(\hat{t})|$ has already been shown to be bounded by (\ref{eq:beta12_mod}) with $\bar{\beta}_u=\beta_1 $ and $\bar{\beta}_y=\beta_2$ from (\ref{eq:betas}), we can write
\begin{align}
	\begin{aligned}
		&\alpha_1(|\Delta x(t)|)\\&\leq \alpha_2(\max_{ \tau \in K_i, \tau < \hat{t}} \beta_2(|\Delta h(\Delta x(\tau))|,\hat{t}-\tau -1 ))e^{-(t-\hat{t})} \\&\oplus \alpha_2(\max_{ 0<\tau < \hat{t}} \beta_1(|\Delta w(\tau)|,\hat{t}-\tau -1 ))e^{-(t-\hat{t})} \\ &\oplus \max_{\hat{t}\leq \tau < t}\alpha_1(\bar{\gamma}_1(| \Delta w(\tau)
	|) \sigma_1(t-\tau-1)), \\ &\phantom{(|\Delta h(\tau)|,\hat{t}-\tau -1 ))e^{-(t-\hat{t})}}\forall t\geq \hat{t}, \ \forall K_i \in K.
	\end{aligned}
\end{align}
Applying (\ref{eq:beta1}) and  (\ref{eq:beta2}) with $\sigma(r)$ selected as $e^{-r}$,  we obtain 
\begin{align}
	\begin{aligned}
		&\alpha_1(|\Delta x(t)|)\\ &\leq \alpha_2(\max_{\substack{ \tau \in K_i,\\ \tau < \hat{t}}} \bar{\gamma}_2(|\Delta h(x(\tau))|)\textcolor{rev}{e^{\tau+2T_{\bar{\beta}}-\hat{t}}}e^{-(t-\hat{t})}\\&\oplus \alpha_2(\max_{ 0<\tau < \hat{t}} \bar{\gamma}_1(|\Delta w(\tau)|) \textcolor{rev}{e^{\tau+2T_{\bar{\beta}}-\hat{t}}}e^{-(t-\hat{t})} \\&\oplus \max_{\hat{t}\leq \tau < t}\alpha_1(\bar{\gamma}_1(| \Delta w(\tau)
	|) \sigma_1(t-\tau-1))\\
		&\leq \alpha_2(\max_{ \substack{\tau \in K_i,\\ \tau < t}} \bar{\gamma}_2(|\Delta h(\Delta x(\tau))|)  \textcolor{rev}{e^{\tau+2T_{\bar{\beta}}-\hat{t}}} \\&\oplus\alpha_2(\max_{ 0<\tau < t} \bar{\gamma}_1(|\Delta w(\tau)|)\textcolor{rev}{e^{\tau+2T_{\bar{\beta}}-\hat{t}}} \\& \oplus \max_{0\leq \tau < t}\alpha_1(\bar{\gamma}_1(| \Delta w(\tau)
	|) \sigma_1(t-\tau-1), \quad  \forall t\geq \hat{t}.
	\end{aligned}
\end{align}
Therefore $|\Delta x(t)|$ is bounded as follows
\begin{align}
	\begin{aligned}
			|\Delta x(t)|&\leq\max_{0\leq \tau< t}\bar{\beta}_u(|\Delta w(\tau)
		|,t-\tau-1)
		\\&	\oplus \max_{ \tau \in K_i, \tau< t}\bar{\beta}_y( |\Delta h(\Delta x(\tau))|,t-\tau-1), \\&  \phantom{h(\Delta x(\tau))|,t-\tau-1),)}  \forall t\geq \hat{t}, \ \forall K_i\in K
	\end{aligned}
\label{eq:bound>hatt} 
\end{align}
with
\begin{subequations}
	\begin{align}
		\begin{split}
			\bar{\beta}_y( s,r)=&\alpha_1^{-1}(\alpha_2( \bar{\gamma}_2(s)  \textcolor{rev}{e^{2T_{\bar{\beta}}-1-r}}))
			\label{eq:thmbetay}
		\end{split}\\
		\begin{split}
			\bar{\beta}_u(s,r)=& \alpha_1^{-1}(\alpha_2( \bar{\gamma}_1(s)\textcolor{rev}{e^{2T_{\bar{\beta}}-1-r}}))\oplus \bar{\gamma}_1(s) \sigma_1(r).
			\label{eq:thmbetau}
		\end{split}
	\end{align}
\end{subequations}
Combining (\ref{eq:tdiISS}), (\ref{eq:PsixwlessT}) and (\ref{eq:bound>hatt})  and noting that we can always choose $\alpha_1$ and $\alpha_2$ such that $\alpha_1^{-1}(\alpha_2(s))\geq s$, we finally obtain
\begin{align}
	\begin{aligned}
			|\Delta x(t)|&\leq\bar{\beta}(|\Delta x_0|,t)\\ &\oplus\max_{0\leq \tau< t}\bar{\beta}_u(|\Delta w(\tau)
		|,t-\tau-1)\\
		&\oplus \max_{ \substack{\tau \in K_i,\\ \tau< t}}\bar{\beta}_y( |\Delta h(\Delta x(\tau))|,t-\tau-1), \\
		&\phantom{(x(\tau))|,t-\tau-1),}\forall t\geq0, \forall K_i\in K.
	\end{aligned}
\end{align}
\end{proof}
To summarize, in Section \ref{sec:sbtdiISS} we extended the results of the previous section to time-discounted \iIOSS/. The already before mentioned Figure~\ref{fig:schem} summarizes the implications that were established in both this section and Section \ref{sec:sbiISS}.

\textcolor{rev}{
\section{Numerical example}
In this paper, we have shown that an \iIOSS/ system is also both sample-based \iIOSS/ and sample-based time-discounted \iIOSS/ if condition (\ref{eq:cond_sampling}) holds.  To illustrate this, a numerical example is presented in this section.
In particular, we numerically determine a bound as in (\ref{eq:cond_sampling}) that holds for a finite time interval and for a set of relevant initial conditions and disturbance sequences. Furthermore, we demonstrate the influence of the sparsity of the set of measurement instances on the bound  (\ref{eq:cond_sampling}).
For this, we consider the simplified two-dimensional model of the hypothalamic–pituitary–thyroid axis from \cite{Yan21}, which models the release of certain hormones in the human body and allows to study thyroid diseases. In the context of such biomedical systems, the measurement of outputs necessitates the collection of blood samples,  which is impractical to realize on a frequent basis. Therefore, the concept of sample-based observability becomes relevant.
The system was discretized using the Euler method with a sampling time of $T=1\mathtt{h}$ resulting in the following system 
\begin{align}
	\begin{aligned}
		x_{TSH}(t+1)&=\frac{p_1T(U-x_{FT4}(t))}{s_1+x_{FT4}(t)}+p_1T\\&+(1-d_1T)x_{TSH}(t)+w_1(t)\\
		x_{FT4}(t+1)&=\frac{T x_{TSH}(t)p_2}{s_2+x_{TSH}(t)}		\\&+(1-d_2T)	x_{FT4}(t)+w_2(t)
	\end{aligned}
	\label{eq:sysex}
\end{align}
with state $x=\begin{pmatrix}x_{TSH}, x_{FT4}\end{pmatrix}$ where $x_{TSH}$ and $x_{FT4}$ denote the concentrations of two thyroid hormones, thyroid-stimulating hormone (TSH) and free thyroxine (FT4), respectively, and with $U$ representing the 
set point of $FT4$. 
We assume $x_{TSH}$ as our measured output, i.e, $h(x)=\begin{pmatrix} 1 &0 \end{pmatrix} x$.
The values of the parameters $p_1,p_2,s_1,s_2$ can be found in \cite{Yan21}. 
We consider that the system is affected by additive disturbances $w_1(t),w_2(t)$ of up to $10 \%$ of the set points of the hormone concentrations. 
 Using the procedure from \cite{Sch23}  i-IOSS can be verified for system (\ref{eq:sysex}).
 }
\par 
\textcolor{rev}{Moreover, we consider a sampling scheme with  $D=\{4,3,5,4,3,5, \ldots\}$ (compare Definition~\ref{def:K}), thus it is assumed that a measurement is available on average every four hours. From the pattern of $D$, notice that the set $K$ is given by $K=\{K_1,K_2,K_3\}$ according to Definition~\ref{def:K}. We simulated the system with all $K_i, \ i=1,2,3$, for different initial states, i.e., $x_{TSH}(0) \in [0.5,4.5]$ and $x_{FT4}(0) \in [12,30]$ and different disturbance sequences over a simulation time of 10 days resulting in 62400 different $\Delta h(\Delta(x(t)))$ trajectories.  We define $\mu:=\{x_{01},x_{02},\boldsymbol{w_1}, \boldsymbol{w_2},K_i\}$ as the initial states, disturbance sequences and measurement sequence used to generate  a single $\Delta h(\Delta(x(t)))$. }
\par 
\textcolor{rev}{ To find a bound as in (\ref{eq:cond_sampling}), we  select linear functions for $\gamma_w, \gamma_h$, i.e., $\gamma_w(s)=c_ws$ and $\gamma_h(s)=c_hs$ with $c_w, c_h \in \mathbb{R}_{\geq 0}$. Our objective now is to show that we can determine constants $c_w$ and $c_h$ such that the condition (\ref{eq:cond_sampling}) holds for all the 62400 simulated trajectories $\Delta h(\Delta(x(t)))$. For this purpose, we propose to fix $t^*=20$ and $c_w=0.2$, such that it only remains to determine a suitable value for $c_h$.
For each $\mu$, a parameter $c_h^{\mu}(t)$ can be calculated at each time instant $t$ that upper bounds the corresponding $|\Delta h(\Delta x(t))|$  as follows
\begin{align}
	\begin{aligned}
		|\Delta h(\Delta x(t))|	&\leq c_{h}^{\mu}(t)\sup_{ \tau \in K_i, \tau < t} | \Delta h(\Delta x(\tau))|\\ &\oplus c_w \sup_{0\leq \tau < t} | \Delta w(\tau)|
	\end{aligned}
\end{align}}
\textcolor{rev}{
with $t\geq t^*$. We define $\bar{c}_h^\mu(t)=\max_{t^*\leq\tau \leq t} c_h^\mu(\tau)$. Figure \ref{fig:K4}  depicts $\bar{c}_h^\mu(t)$ for some exemplary selections of $\mu$. To obtain $c_h$ that holds for all  62400 simulated $\mu$ over the complete simulation time of 10 days, we determine $c_h=\max_{\mu}\bar{c}_h^\mu(10 d)$, resulting in $c_h=3.02$. The corresponding $\bar{c}_h^{\mu_{\mathtt{max}}}(t)$ with $\mu_{\mathtt{max}}=\argmax_{\mu}c_h^\mu(10 d)$  is depicted by a red dotted line in Figure~\ref{fig:K4}.}\textcolor{rev}{
\begin{figure}[!t]
	\setlength\figurewidth{0.9\columnwidth} 
	\flushleft{\input{figures/K4}}
	\caption{$\bar{c}_h^\mu(t)$ for different $\mu$ (blue dashdotted), and  $\bar{c}_h^{\mu_{\mathtt{max}}}(t)$ (red dotted).}
\label{fig:K4}
\end{figure}}
\par \textcolor{rev}{Furthermore, to illustrate the effect of the sparsity of a sampling  scheme, we repeated the simulations for different $K$, namely for sampling schemes where on average every 2nd, 10th or 14th measurement is available. In order to identify the impact of sparsity on the bound (\ref{eq:cond_sampling}), we maintained the same fixed values for $c_w$ and $t^*$. Figure~\ref{fig:K} shows $\bar{c}_h^{\mu_{\mathtt{max}}}(t)$ for the }\textcolor{rev}{different sampling schemes. It can be observed that a reduction in the number of samples leads to an increase in the value }\textcolor{rev}{of $c_h$, consequently resulting in a larger bound for (\ref{eq:cond_sampling}).}\textcolor{rev}{ 
\begin{figure}[!t]
\setlength\figurewidth{0.9\columnwidth} 
\flushleft{\input{figures/K_sparsityx}}
\caption{$\bar{c}_h^{\mu_{\mathtt{max}}}(t)$ for different sampling schemes that consider on average every 2nd ($K2$), 4th ($K4$), 10th ($K10$),  14th ($K14$) measurement. }
\label{fig:K}
\end{figure}
}\par\textcolor{rev}{ In conclusion, we have have numerically verified the bound (\ref{eq:cond_sampling}) by calculating a function  $\gamma_h$  for a set of disturbance sequences and initial conditions relevant to the application, over a finite time interval, and demonstrated the impact of the frequency of measured outputs. This then lets us conclude that the system is both sample-based i-IOSS and sample-based time-discounted i-IOSS w.r.t. the different considered choices of $K$, according to Theorems~\ref{thm1} and \ref{thm1_td}, respectively.
}

\section{Conclusion}
\label{sec:con}
In this paper, we studied detectability conditions for discrete-time nonlinear systems in case of irregular measurement sequences. A sample-based version of \iIOSS/ was formulated, that considers in the output-dependent term of the \iIOSS/ bound only the limited output information from the irregular sampling of the system. Furthermore, we explored a time-discounted version of  sample-based \iIOSS/. We provided conditions for an \iIOSS/ system to be sample-based \iIOSS/ and  \textcolor{rev}{sample-based time-discounted \iIOSS/}. 
\par Future research will focus on exploiting the presented sample-based detectability conditions to design  sample-based nonlinear estimators, in particular   moving horizon estimators. This will enable state estimation and, in turn, control in applications where only few and irregular measurements are available.
\textcolor{rev}{	Moreover, we envision further potential practical uses for the condition (\ref{eq:cond_sampling}) for example in the context of event-triggered state estimation, where the triggering mechanism may use (\ref{eq:cond_sampling}) to determine which samples to send to the estimator, such that the system states can be reconstructed.
	Besides,  it would be a valuable direction for future research to study sampling schemes for specific classes of nonlinear systems to satisfy (\ref{eq:cond_sampling}).
	}

\appendices

\begin{IEEEbiography}[{\includegraphics[width=1in,height=1.25in,clip,keepaspectratio]{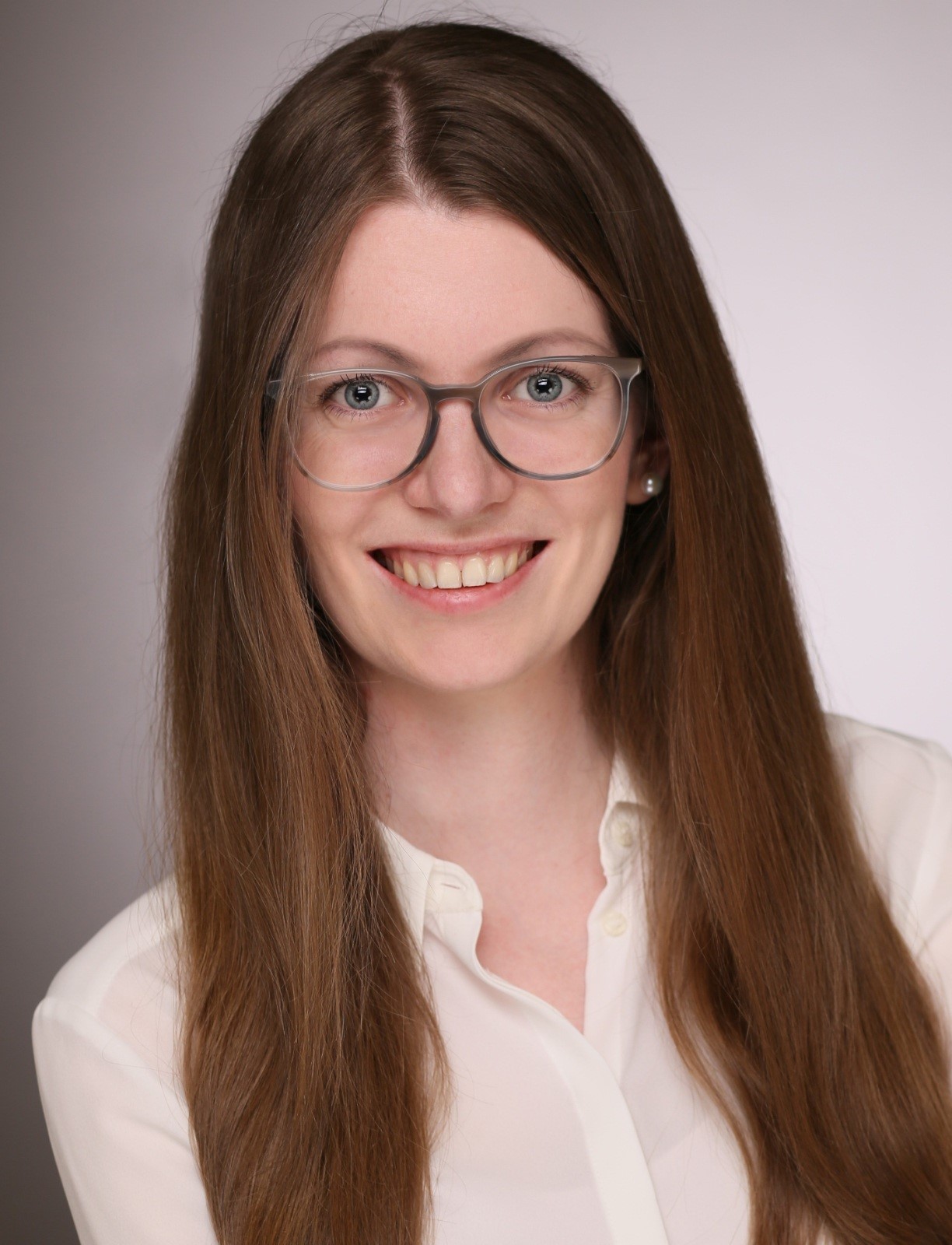}}]{Isabelle Krauss} received her Master degree
	in Automation Engineering from RWTH Aachen University, Germany, in 2020. Since then, she
	has been a research assistant at the Institute
	of Automatic Control, Leibniz University Hannover, Germany, where she is currently working on her
	Ph.D. under the supervision of Prof. Matthias
	A. Müller. Her research interests are in the area of observability and state estimation with focus on nonlinear systems. 
\end{IEEEbiography}
\begin{IEEEbiography}[{\includegraphics[width=1in,height=1.25in,clip,keepaspectratio]{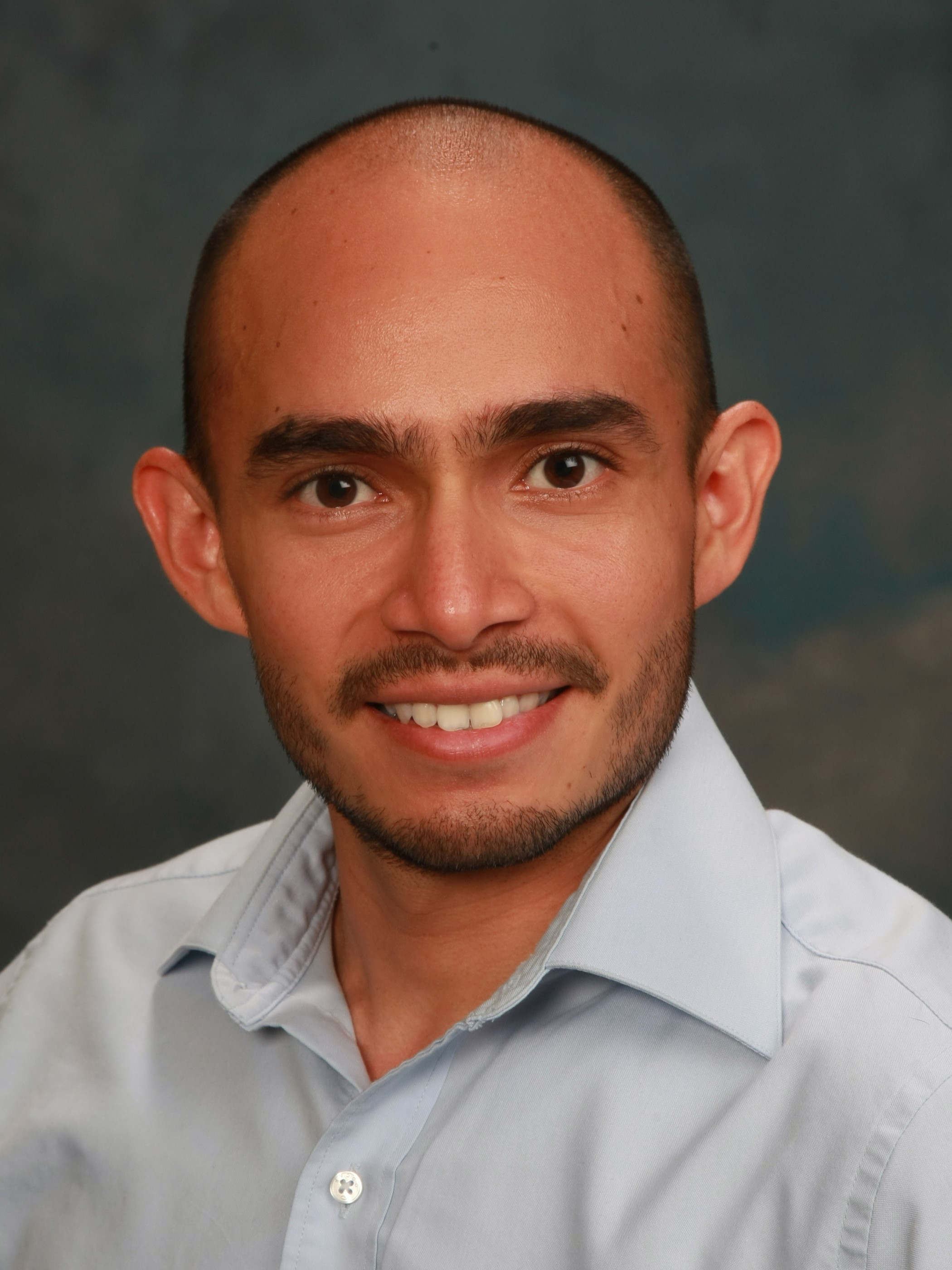}}]{Victor G. Lopez} (Member, IEEE)
	received his B.Sc. degree in Communications and Electronics Engineering from the Universidad Autonoma de Campeche, in Campeche, Mexico, in 2010, the M.Sc. degree in Electrical Engineering from the Research and Advanced Studies Center (Cinvestav), in Guadalajara, Mexico, in 2013, and his PhD degree in Electrical Engineering from the University of Texas at Arlington, Texas, USA, in 2019. In 2015 Victor was a Lecturer at the Western Technological Institute of Superior Studies (ITESO) in Guadalajara, Mexico. From August 2019 to June 2020, he was a postdoctoral researcher at the University of Texas at Arlington Research Institute and an Adjunct Professor in the Electrical Engineering department at UTA. Victor is currently a postdoctoral researcher at the Institute of Automatic Control, Leibniz University Hanover, in Hanover, Germany. His research interest include cyber-physical systems, reinforcement learning, game theory, distributed control and robust control. 
\end{IEEEbiography}
\begin{IEEEbiography}[{\includegraphics[width=1in,height=1.25in,clip,keepaspectratio]{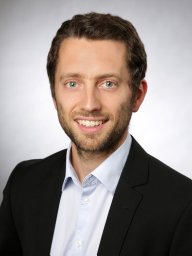}}]{Matthias A. Müller} (Senior Member, IEEE) received a Diploma degree in engineering cybernetics from the University of Stuttgart, Germany, an M.Sc. in electrical and computer engineering from the University of Illinois at Urbana-Champaign, US (both in 2009), and a Ph.D. from the University of Stuttgart in 2014. Since 2019, he is Director of the Institute of Automatic Control and Full Professor at the Leibniz University Hannover, Germany.
His research interests include nonlinear control and estimation, model predictive control, and data- and learning-based control, with application in different fields including biomedical engineering and robotics. He has received various awards for his work, including the 2015 European Systems \& Control PhD Thesis Award, the inaugural Brockett-Willems Outstanding Paper Award for the best paper published in Systems \& Control Letters in the period 2014-2018, an ERC starting grant in 2020, the IEEE CSS George S. Axelby Outstanding Paper Award 2022, and the Journal of Process Control Paper Award 2023. He serves as an associate editor for Automatica and as an editor of the International Journal of Robust and Nonlinear Control.
\end{IEEEbiography}

\end{document}

%% file: figures/tp_mod.tex
%
%
\definecolor{mycolor1}{RGB} {0,80,155}%
\begin{tikzpicture}
	
	\begin{axis}[%
		width=0.98\figurewidth,
		height=0.75\figurewidth,
		at={(0\figurewidth, 0\figurewidth)},
		scale only axis,
		xmin=0,
		xmax=150,
		ymin=-400,
		ymax=300,
		xlabel={\textcolor{rev}{$t$}},
		ylabel={$\Delta y(t)$},
		axis background/.style={fill=white},
		x tick label style={font=\small},
		y tick label style={font=\small},
		xmajorgrids,
		ymajorgrids,
		legend style={at={(0.03,0.97)}, anchor=north west, legend cell align=left, align=left, draw=black, font=\small}
		]
		\addplot[const plot,color=mycolor1] table[row sep=crcr] {%
			0	1\\
			1	1\\
			2	0.6175\\
			3	-0.0327500000000001\\
			4	-0.72411875\\
			5	-1.19555\\
			6	-1.248576453125\\
			7	-0.828397095312499\\
			8	-0.0566910515234367\\
			9	0.800365068085938\\
			10	1.42198867902022\\
			11	1.55098556813134\\
			12	1.09737272078389\\
			13	0.186591747830441\\
			14	-0.870699998936812\\
			15	-1.68217556521903\\
			16	-1.91716571202326\\
			17	-1.43822666108993\\
			18	-0.369653440587712\\
			19	0.928469511630743\\
			20	1.97854801920846\\
			21	2.3584633863141\\
			22	1.86760952594082\\
			23	0.621899578414373\\
			24	-0.964458028526502\\
			25	-2.31278494212861\\
			26	-2.8877085857387\\
			27	-2.40551489590157\\
			28	-0.963430778970524\\
			29	0.966137550563559\\
			30	2.68534765419364\\
			31	3.51924711364414\\
			32	3.07583125753042\\
			33	1.41932813728193\\
			34	-0.916729502897395\\
			35	-3.09486286353326\\
			36	-4.26890718112012\\
			37	-3.90695315812944\\
			38	-2.02072834525752\\
			39	0.794038606737332\\
			40	3.53730406519473\\
			41	5.15387011903788\\
			42	4.9324475517911\\
			43	2.80609643418637\\
			44	-0.569010536697044\\
			45	-4.00491730239172\\
			46	-6.19240550809137\\
			47	-6.19176638391629\\
			48	-3.82272389014879\\
			49	0.203956497336443\\
			50	4.48482465655802\\
			51	7.40341900778193\\
			52	7.73098962250523\\
			53	5.12848128233495\\
			54	0.349621913607496\\
			55	-4.95722373499483\\
			56	-8.80574607097133\\
			57	-9.60357362751936\\
			58	-6.79385504495644\\
			59	-1.15368512463624\\
			60	5.39308337428374\\
			61	10.4171058837011\\
			62	11.8710672496297\\
			63	8.90429720526406\\
			64	2.28687495122476\\
			65	-5.75121430761626\\
			66	-12.2526064576484\\
			67	-14.6037414900077\\
			68	-11.5629140426087\\
			69	-3.84840370950145\\
			70	5.97456814497148\\
			71	14.3226628619868\\
			72	17.881056848446\\
			73	14.8935140942574\\
			74	5.96272992179487\\
			75	-5.98558813998241\\
			76	-16.630154978313\\
			77	-21.7918643016012\\
			78	-19.0440265486982\\
			79	-8.78515202630361\\
			80	5.68040029424966\\
			81	19.1666075686981\\
			82	26.4341995482615\\
			83	24.1902865389288\\
			84	12.508466105186\\
			85	-4.92159279957431\\
			86	-21.9071223181401\\
			87	-31.9144837352991\\
			88	-30.5401624406217\\
			89	-17.3708475055956\\
			90	3.52928508246039\\
			91	24.8037270649899\\
			92	38.3458849087195\\
			93	38.3379697969716\\
			94	23.6651282411628\\
			95	-1.27013429524497\\
			96	-27.7767296229752\\
			97	-45.8455199844551\\
			98	-47.8690741567031\\
			99	-31.7496506832226\\
			100	-2.15613338684727\\
			101	30.7035701069481\\
			102	54.5300835730739\\
			103	59.4645274334543\\
			104	42.0608811690199\\
			105	7.13314704061945\\
			106	-33.4045538964109\\
			107	-64.5093732953691\\
			108	-73.5055050092627\\
			109	-55.1279619235095\\
			110	-14.1478260974392\\
			111	35.6247144165523\\
			112	75.8770362586169\\
			113	90.4272082837309\\
			114	71.5893623323897\\
			115	23.8144629979238\\
			116	-37.0108976283414\\
			117	-88.6976821634329\\
			118	-110.721762995156\\
			119	-92.2117561498496\\
			120	-36.9036770124876\\
			121	37.0829751109832\\
			122	102.989288054689\\
			123	134.939469135333\\
			124	117.911193210864\\
			125	54.3770531194715\\
			126	-35.1978763476589\\
			127	-118.699549792848\\
			128	-163.687533501501\\
			129	-149.776544301793\\
			130	-77.4283702976742\\
			131	30.5048797006453\\
			132	135.674506338329\\
			133	197.625128499211\\
			134	189.095065337418\\
			135	107.532409473214\\
			136	-21.8903121232911\\
			137	-153.617363864349\\
			138	-237.45325569593\\
			139	-237.379738299924\\
			140	-146.502405819026\\
			141	7.90947681732331\\
			142	172.034964888546\\
			143	283.897431655775\\
			144	296.397784322967\\
			145	196.557263581667\\
			146	13.2967465592227\\
			147	-190.168768676476\\
			148	-337.680634900368\\
			149	-368.19938723834\\
			150	-260.39967102553\\
		};
		\addlegendentry{Not sampled}
		\addplot[color=red, mark=x, mark size=2.5, only marks] table[row sep=crcr] {%
		6	-1.248576453125\\
		20	1.97854801920846\\
		40	3.53730406519473\\
		42	4.9324475517911\\
		57	-9.60357362751936\\
		71	14.3226628619868\\
		87	-31.9144837352991\\
		90	3.52928508246039\\
		99	-31.7496506832226\\
		105	7.13314704061945\\
		115	23.8144629979238\\
		136	-21.8903121232911\\
		146	13.2967465592227\\
	};
		\addlegendentry{Sampled }
	\end{axis}

\end{tikzpicture}%

%% file: figures/overview.tex
\begin{tikzpicture}
	\usetikzlibrary{arrows}
	\usetikzlibrary{shapes.geometric}
	\begin{small}
	\tikzstyle{block} = [draw, minimum width=3cm, minimum height=1.2cm,  rounded corners=.4cm, align = center]
	[shorten -2pt,shorten -2pt,-implies]
	\node[block] (ioss)  at (3,8) {i-IOSS};
	\node[block] (ioss11)  at (-2,8) {i-IOSS\\ + (11) holds for all \\$ (x_{01},x_{02},\boldsymbol{w_1},\boldsymbol{w_2})\in \Psi$};
	\node[block] (sb) at (-2,1.5) {sample-based\\ i-IOSS};
	\node[block] (td) at (3,1.5) {sample-based\\ time-discounted\\ i-IOSS};

	\draw[shorten >=0.1cm,shorten <=0.1cm,-implies,double equal sign distance] ([yshift=1.4mm]td.west) -- ([yshift=1.4mm]sb.east) node [above,midway] { Lemma 1};
	\draw[shorten >=0.1cm,shorten <=0.1cm,-implies,double equal sign distance] ([yshift=-1.4mm]sb.east) -- ([yshift=-1.4mm]td.west) node [below,midway,text width=1.9cm] {Theorem 4 [Ass. 1,2,3, $\boldsymbol{w_1},\boldsymbol{w_2}\in W$]}; 
	\draw[shorten >=0.1cm,shorten <=0.1cm,-implies,double equal sign distance] ([xshift=11.2mm]sb.north) -- ([yshift=-4.2mm]ioss.west) ;
	\draw[shorten >=0.1cm,shorten <=0.1cm,-implies,double equal sign distance] ([xshift=-11.2mm]ioss.south) -- ([yshift=4.2mm]sb.east) node  [right, pos=0.65,text width=1.9cm]{\phantom{..}Theorem 1 [Ass. 1, (11)]};
	\draw[shorten >=0.1cm,shorten <=0.1cm,-implies,double equal sign distance] ([xshift=1.8mm]ioss.south) -- ([xshift=1.8mm]td.north)node  [right, pos=0.5,text width=1.74cm] {Theorem 3 [Ass. 1, (11)]};
	\draw[shorten >=0.1cm,shorten <=0.1cm,-implies,double equal sign distance] ([xshift=-1.8mm]td.north) -- ([xshift=-1.8mm]ioss.south);
	\draw[shorten >=0.1cm,shorten <=0.1cm,-implies,double equal sign distance] ([xshift=1.8mm]ioss11.south) -- ([xshift=1.8mm]sb.north)node [right, pos=0.3,text width=2.241cm] {Corollary~1 \mbox{[Ass. 1,} $\boldsymbol{w_1},\boldsymbol{w_2}\in W$]};
	\draw[shorten >=0.1cm,shorten <=0.1cm,-implies,double equal sign distance] ([xshift=-1.8mm]sb.north) -- ([xshift=-1.8mm]ioss11.south) node [left, pos=0.3,text width=1.41cm] {Theorem 2 [Ass. 1, 2]};
	\draw[shorten >=0.1cm,shorten <=0.1cm,-implies,double equal sign distance] (ioss11.east) -- (ioss.west);
\end{small}
\end{tikzpicture}

%% file: figures/K4.tex
%
%
\definecolor{mycolor1}{rgb}{0.00000,0.60000,0.90000}
\definecolor{mycolor2}{rgb}{0.00000,0.60000,0.90000}
\definecolor{mycolor3}{rgb}{0.00000,0.60000,0.90000}
\begin{tikzpicture}

\begin{axis}[%
	width=0.96\figurewidth,
height=0.85\figurewidth,
at={(0\figurewidth, 0\figurewidth)},
scale only axis,
xmin=20,
xmax=240,
xlabel style={font=\color{white!15!black}},
xlabel={$t$ in $\mathtt{h}$},
ymin=0,
ymax=3.5,
ylabel style={font=\color{white!15!black},at={(axis description cs:0.05,0.5)}},
ylabel={${\bar{c}}_{h}^\mu(t)$},
xtick={20,50,100,150,200},
axis background/.style={fill=white}
]
\addplot [color=red, dotted, line width=2.3pt, forget plot]
  table[row sep=crcr]{%
0	0\\
1	0\\
2	0\\
3	0\\
4	0\\
5	0\\
6	0\\
7	0\\
8	0\\
9	0\\
10	0\\
11	0\\
12	0\\
13	0\\
14	0\\
15	0\\
16	0\\
17	0\\
18	0\\
19	0\\
20	0\\
21	0\\
22	0\\
23	0\\
24	0\\
25	0\\
26	0\\
27	0\\
28	0\\
29	0\\
30	0\\
31	0\\
32	0\\
33	0\\
34	0\\
35	0\\
36	0\\
37	0\\
38	0\\
39	0\\
40	0\\
41	0\\
42	0\\
43	0\\
44	0\\
45	0\\
46	0\\
47	0\\
48	0\\
49	0\\
50	0\\
51	0\\
52	0\\
53	0\\
54	0\\
55	0\\
56	0.0258319499773168\\
57	0.1351911657352\\
58	0.1351911657352\\
59	0.1351911657352\\
60	0.1351911657352\\
61	0.1351911657352\\
62	0.1351911657352\\
63	0.1351911657352\\
64	0.1351911657352\\
65	0.1351911657352\\
66	0.1351911657352\\
67	0.1351911657352\\
68	0.1351911657352\\
69	0.1351911657352\\
70	0.1351911657352\\
71	0.1351911657352\\
72	0.1351911657352\\
73	0.1351911657352\\
74	0.1351911657352\\
75	0.1351911657352\\
76	0.1351911657352\\
77	0.1351911657352\\
78	0.1351911657352\\
79	0.1351911657352\\
80	3.02197958138586\\
81	3.02197958138586\\
82	3.02197958138586\\
83	3.02197958138586\\
84	3.02197958138586\\
85	3.02197958138586\\
86	3.02197958138586\\
87	3.02197958138586\\
88	3.02197958138586\\
89	3.02197958138586\\
90	3.02197958138586\\
91	3.02197958138586\\
92	3.02197958138586\\
93	3.02197958138586\\
94	3.02197958138586\\
95	3.02197958138586\\
96	3.02197958138586\\
97	3.02197958138586\\
98	3.02197958138586\\
99	3.02197958138586\\
100	3.02197958138586\\
101	3.02197958138586\\
102	3.02197958138586\\
103	3.02197958138586\\
104	3.02197958138586\\
105	3.02197958138586\\
106	3.02197958138586\\
107	3.02197958138586\\
108	3.02197958138586\\
109	3.02197958138586\\
110	3.02197958138586\\
111	3.02197958138586\\
112	3.02197958138586\\
113	3.02197958138586\\
114	3.02197958138586\\
115	3.02197958138586\\
116	3.02197958138586\\
117	3.02197958138586\\
118	3.02197958138586\\
119	3.02197958138586\\
120	3.02197958138586\\
121	3.02197958138586\\
122	3.02197958138586\\
123	3.02197958138586\\
124	3.02197958138586\\
125	3.02197958138586\\
126	3.02197958138586\\
127	3.02197958138586\\
128	3.02197958138586\\
129	3.02197958138586\\
130	3.02197958138586\\
131	3.02197958138586\\
132	3.02197958138586\\
133	3.02197958138586\\
134	3.02197958138586\\
135	3.02197958138586\\
136	3.02197958138586\\
137	3.02197958138586\\
138	3.02197958138586\\
139	3.02197958138586\\
140	3.02197958138586\\
141	3.02197958138586\\
142	3.02197958138586\\
143	3.02197958138586\\
144	3.02197958138586\\
145	3.02197958138586\\
146	3.02197958138586\\
147	3.02197958138586\\
148	3.02197958138586\\
149	3.02197958138586\\
150	3.02197958138586\\
151	3.02197958138586\\
152	3.02197958138586\\
153	3.02197958138586\\
154	3.02197958138586\\
155	3.02197958138586\\
156	3.02197958138586\\
157	3.02197958138586\\
158	3.02197958138586\\
159	3.02197958138586\\
160	3.02197958138586\\
161	3.02197958138586\\
162	3.02197958138586\\
163	3.02197958138586\\
164	3.02197958138586\\
165	3.02197958138586\\
166	3.02197958138586\\
167	3.02197958138586\\
168	3.02197958138586\\
169	3.02197958138586\\
170	3.02197958138586\\
171	3.02197958138586\\
172	3.02197958138586\\
173	3.02197958138586\\
174	3.02197958138586\\
175	3.02197958138586\\
176	3.02197958138586\\
177	3.02197958138586\\
178	3.02197958138586\\
179	3.02197958138586\\
180	3.02197958138586\\
181	3.02197958138586\\
182	3.02197958138586\\
183	3.02197958138586\\
184	3.02197958138586\\
185	3.02197958138586\\
186	3.02197958138586\\
187	3.02197958138586\\
188	3.02197958138586\\
189	3.02197958138586\\
190	3.02197958138586\\
191	3.02197958138586\\
192	3.02197958138586\\
193	3.02197958138586\\
194	3.02197958138586\\
195	3.02197958138586\\
196	3.02197958138586\\
197	3.02197958138586\\
198	3.02197958138586\\
199	3.02197958138586\\
200	3.02197958138586\\
201	3.02197958138586\\
202	3.02197958138586\\
203	3.02197958138586\\
204	3.02197958138586\\
205	3.02197958138586\\
206	3.02197958138586\\
207	3.02197958138586\\
208	3.02197958138586\\
209	3.02197958138586\\
210	3.02197958138586\\
211	3.02197958138586\\
212	3.02197958138586\\
213	3.02197958138586\\
214	3.02197958138586\\
215	3.02197958138586\\
216	3.02197958138586\\
217	3.02197958138586\\
218	3.02197958138586\\
219	3.02197958138586\\
220	3.02197958138586\\
221	3.02197958138586\\
222	3.02197958138586\\
223	3.02197958138586\\
224	3.02197958138586\\
225	3.02197958138586\\
226	3.02197958138586\\
227	3.02197958138586\\
228	3.02197958138586\\
229	3.02197958138586\\
230	3.02197958138586\\
231	3.02197958138586\\
232	3.02197958138586\\
233	3.02197958138586\\
234	3.02197958138586\\
235	3.02197958138586\\
236	3.02197958138586\\
237	3.02197958138586\\
238	3.02197958138586\\
239	3.02197958138586\\
240	3.02197958138586\\
};
\addplot [color=mycolor3, dashdotted, line width=1.5pt, forget plot]
  table[row sep=crcr]{%
0	0\\
1	0\\
2	0\\
3	0\\
4	0\\
5	0\\
6	0\\
7	0\\
8	0\\
9	0\\
10	0\\
11	0\\
12	0\\
13	0\\
14	0\\
15	0\\
16	0\\
17	0\\
18	0\\
19	0\\
20	0.0117235443823615\\
21	0.0117235443823615\\
22	0.0264199406063688\\
23	0.0264199406063688\\
24	0.0264199406063688\\
25	0.0264199406063688\\
26	0.0264199406063688\\
27	0.0264199406063688\\
28	0.0264199406063688\\
29	0.141740288658498\\
30	0.141740288658498\\
31	0.156725173574049\\
32	0.161414890011988\\
33	0.161414890011988\\
34	0.161414890011988\\
35	0.161414890011988\\
36	0.161414890011988\\
37	0.161414890011988\\
38	0.199425285926684\\
39	0.209800681690823\\
40	0.257521311636958\\
41	0.257521311636958\\
42	0.257521311636958\\
43	0.257521311636958\\
44	0.257521311636958\\
45	0.257521311636958\\
46	0.257521311636958\\
47	0.257521311636958\\
48	0.33088489413077\\
49	0.33088489413077\\
50	0.33088489413077\\
51	0.33088489413077\\
52	0.33088489413077\\
53	0.446116707328714\\
54	0.446116707328714\\
55	0.446116707328714\\
56	0.446116707328714\\
57	0.451266491318668\\
58	0.451266491318668\\
59	0.451266491318668\\
60	0.451266491318668\\
61	0.451266491318668\\
62	0.451266491318668\\
63	0.451266491318668\\
64	0.451266491318668\\
65	0.451266491318668\\
66	0.451266491318668\\
67	0.451266491318668\\
68	0.451266491318668\\
69	0.451266491318668\\
70	0.451266491318668\\
71	0.451266491318668\\
72	0.451266491318668\\
73	0.451266491318668\\
74	0.451266491318668\\
75	0.451266491318668\\
76	0.451266491318668\\
77	0.451266491318668\\
78	0.451266491318668\\
79	0.451266491318668\\
80	0.451266491318668\\
81	0.451266491318668\\
82	0.451266491318668\\
83	0.451266491318668\\
84	0.451266491318668\\
85	0.451266491318668\\
86	0.485958750726835\\
87	0.485958750726835\\
88	0.485958750726835\\
89	0.485958750726835\\
90	0.485958750726835\\
91	0.485958750726835\\
92	0.485958750726835\\
93	0.485958750726835\\
94	0.485958750726835\\
95	0.485958750726835\\
96	0.485958750726835\\
97	0.485958750726835\\
98	0.485958750726835\\
99	0.485958750726835\\
100	0.485958750726835\\
101	0.485958750726835\\
102	0.485958750726835\\
103	0.485958750726835\\
104	0.485958750726835\\
105	0.485958750726835\\
106	0.485958750726835\\
107	0.485958750726835\\
108	0.485958750726835\\
109	0.485958750726835\\
110	0.527448215332385\\
111	0.527448215332385\\
112	0.527448215332385\\
113	0.527448215332385\\
114	0.527448215332385\\
115	0.527448215332385\\
116	0.527448215332385\\
117	0.527448215332385\\
118	0.527448215332385\\
119	0.527448215332385\\
120	0.527448215332385\\
121	0.527448215332385\\
122	0.527448215332385\\
123	0.527448215332385\\
124	0.527448215332385\\
125	0.527448215332385\\
126	0.527448215332385\\
127	0.527448215332385\\
128	0.527448215332385\\
129	0.575914264799899\\
130	0.575914264799899\\
131	0.575914264799899\\
132	0.575914264799899\\
133	0.575914264799899\\
134	0.575914264799899\\
135	0.575914264799899\\
136	0.575914264799899\\
137	0.575914264799899\\
138	0.575914264799899\\
139	0.575914264799899\\
140	0.575914264799899\\
141	0.575914264799899\\
142	0.575914264799899\\
143	0.575914264799899\\
144	0.575914264799899\\
145	0.575914264799899\\
146	0.575914264799899\\
147	0.575914264799899\\
148	0.575914264799899\\
149	0.575914264799899\\
150	0.575914264799899\\
151	0.898147929132434\\
152	0.898147929132434\\
153	0.898147929132434\\
154	0.898147929132434\\
155	0.898147929132434\\
156	0.898147929132434\\
157	0.898147929132434\\
158	0.898147929132434\\
159	0.898147929132434\\
160	0.898147929132434\\
161	0.898147929132434\\
162	0.898147929132434\\
163	0.898147929132434\\
164	0.898147929132434\\
165	0.898147929132434\\
166	0.898147929132434\\
167	0.898147929132434\\
168	0.898147929132434\\
169	0.898147929132434\\
170	0.898147929132434\\
171	0.898147929132434\\
172	0.898147929132434\\
173	0.898147929132434\\
174	0.898147929132434\\
175	0.898147929132434\\
176	0.898147929132434\\
177	0.898147929132434\\
178	0.898147929132434\\
179	0.898147929132434\\
180	0.898147929132434\\
181	0.898147929132434\\
182	0.898147929132434\\
183	0.898147929132434\\
184	0.898147929132434\\
185	0.898147929132434\\
186	0.898147929132434\\
187	0.898147929132434\\
188	0.898147929132434\\
189	0.898147929132434\\
190	0.898147929132434\\
191	0.898147929132434\\
192	0.898147929132434\\
193	0.898147929132434\\
194	0.898147929132434\\
195	0.898147929132434\\
196	0.898147929132434\\
197	0.898147929132434\\
198	0.898147929132434\\
199	0.898147929132434\\
200	0.898147929132434\\
201	0.898147929132434\\
202	0.898147929132434\\
203	0.898147929132434\\
204	0.898147929132434\\
205	0.898147929132434\\
206	0.898147929132434\\
207	0.898147929132434\\
208	0.898147929132434\\
209	0.898147929132434\\
210	0.898147929132434\\
211	0.898147929132434\\
212	0.898147929132434\\
213	0.898147929132434\\
214	0.898147929132434\\
215	0.898147929132434\\
216	0.898147929132434\\
217	0.898147929132434\\
218	0.898147929132434\\
219	0.898147929132434\\
220	0.898147929132434\\
221	0.898147929132434\\
222	0.898147929132434\\
223	0.898147929132434\\
224	0.898147929132434\\
225	0.898147929132434\\
226	0.898147929132434\\
227	0.898147929132434\\
228	0.898147929132434\\
229	0.898147929132434\\
230	0.898147929132434\\
231	0.898147929132434\\
232	0.898147929132434\\
233	0.898147929132434\\
234	0.898147929132434\\
235	0.898147929132434\\
236	0.898147929132434\\
237	0.898147929132434\\
238	0.898147929132434\\
239	0.898147929132434\\
240	0.898147929132434\\
};
\addplot [color=mycolor1, dashdotted, line width=1.5pt, forget plot]
  table[row sep=crcr]{%
0	0\\
1	0\\
2	0\\
3	0\\
4	0\\
5	0\\
6	0\\
7	0\\
8	0\\
9	0\\
10	0\\
11	0\\
12	0\\
13	0\\
14	0\\
15	0\\
16	0\\
17	0\\
18	0\\
19	0\\
20	0.347903957011725\\
21	0.381627159202536\\
22	0.381627159202536\\
23	0.381627159202536\\
24	0.381627159202536\\
25	0.418507648625052\\
26	0.418507648625052\\
27	0.418507648625052\\
28	0.418507648625052\\
29	0.418507648625052\\
30	0.418507648625052\\
31	0.418507648625052\\
32	0.418507648625052\\
33	0.418507648625052\\
34	0.418507648625052\\
35	0.418507648625052\\
36	0.418507648625052\\
37	0.418507648625052\\
38	0.418507648625052\\
39	0.418507648625052\\
40	0.418507648625052\\
41	0.418507648625052\\
42	0.418507648625052\\
43	0.418507648625052\\
44	0.418507648625052\\
45	0.418507648625052\\
46	0.418507648625052\\
47	0.437403243642177\\
48	0.437403243642177\\
49	0.437403243642177\\
50	0.437403243642177\\
51	0.437403243642177\\
52	0.437403243642177\\
53	0.437403243642177\\
54	0.437403243642177\\
55	0.437403243642177\\
56	0.437403243642177\\
57	0.437403243642177\\
58	0.437403243642177\\
59	0.437403243642177\\
60	0.437403243642177\\
61	0.437403243642177\\
62	0.437403243642177\\
63	0.437403243642177\\
64	0.437403243642177\\
65	0.437403243642177\\
66	0.437403243642177\\
67	0.437403243642177\\
68	0.437403243642177\\
69	0.437403243642177\\
70	0.437403243642177\\
71	0.437403243642177\\
72	0.437403243642177\\
73	0.437403243642177\\
74	0.437403243642177\\
75	0.437403243642177\\
76	0.437403243642177\\
77	0.437403243642177\\
78	0.508731070651059\\
79	0.508843614558508\\
80	0.544628094702772\\
81	0.544628094702772\\
82	0.544628094702772\\
83	0.544628094702772\\
84	0.548464450812394\\
85	0.548464450812394\\
86	0.548464450812394\\
87	0.548464450812394\\
88	0.548464450812394\\
89	0.548464450812394\\
90	0.548464450812394\\
91	0.548464450812394\\
92	0.548464450812394\\
93	0.548464450812394\\
94	0.568682644779224\\
95	0.568682644779224\\
96	0.568682644779224\\
97	0.568682644779224\\
98	0.568682644779224\\
99	0.568682644779224\\
100	0.568682644779224\\
101	0.568682644779224\\
102	0.568682644779224\\
103	0.568682644779224\\
104	0.568682644779224\\
105	0.568682644779224\\
106	0.568682644779224\\
107	0.568682644779224\\
108	0.568682644779224\\
109	0.568682644779224\\
110	1.22737371398832\\
111	1.22737371398832\\
112	1.22737371398832\\
113	1.22737371398832\\
114	1.22737371398832\\
115	1.22737371398832\\
116	1.22737371398832\\
117	1.22737371398832\\
118	1.22737371398832\\
119	1.22737371398832\\
120	1.22737371398832\\
121	1.22737371398832\\
122	1.22737371398832\\
123	1.22737371398832\\
124	1.22737371398832\\
125	1.22737371398832\\
126	1.22737371398832\\
127	1.22737371398832\\
128	1.22737371398832\\
129	1.22737371398832\\
130	1.22737371398832\\
131	1.22737371398832\\
132	1.22737371398832\\
133	1.22737371398832\\
134	1.22737371398832\\
135	1.22737371398832\\
136	1.22737371398832\\
137	1.22737371398832\\
138	1.22737371398832\\
139	1.22737371398832\\
140	1.22737371398832\\
141	1.22737371398832\\
142	1.22737371398832\\
143	1.22737371398832\\
144	1.22737371398832\\
145	1.22737371398832\\
146	1.22737371398832\\
147	1.22737371398832\\
148	1.22737371398832\\
149	1.22737371398832\\
150	1.22737371398832\\
151	1.22737371398832\\
152	1.22737371398832\\
153	1.22737371398832\\
154	1.22737371398832\\
155	1.22737371398832\\
156	1.22737371398832\\
157	1.22737371398832\\
158	1.22737371398832\\
159	1.22737371398832\\
160	1.22737371398832\\
161	1.22737371398832\\
162	1.22737371398832\\
163	1.22737371398832\\
164	1.22737371398832\\
165	1.22737371398832\\
166	1.22737371398832\\
167	1.22737371398832\\
168	1.22737371398832\\
169	1.22737371398832\\
170	1.22737371398832\\
171	1.22737371398832\\
172	1.22737371398832\\
173	1.22737371398832\\
174	1.22737371398832\\
175	1.22737371398832\\
176	1.22737371398832\\
177	1.22737371398832\\
178	1.22737371398832\\
179	1.22737371398832\\
180	1.22737371398832\\
181	1.22737371398832\\
182	1.22737371398832\\
183	1.22737371398832\\
184	1.22737371398832\\
185	1.22737371398832\\
186	1.22737371398832\\
187	1.22737371398832\\
188	1.22737371398832\\
189	1.22737371398832\\
190	1.22737371398832\\
191	1.22737371398832\\
192	1.22737371398832\\
193	1.22737371398832\\
194	1.22737371398832\\
195	1.22737371398832\\
196	1.22737371398832\\
197	1.22737371398832\\
198	1.22737371398832\\
199	1.22737371398832\\
200	1.22737371398832\\
201	1.22737371398832\\
202	1.22737371398832\\
203	1.22737371398832\\
204	1.22737371398832\\
205	1.22737371398832\\
206	1.22737371398832\\
207	1.22737371398832\\
208	1.22737371398832\\
209	1.22737371398832\\
210	1.22737371398832\\
211	1.22737371398832\\
212	1.22737371398832\\
213	1.22737371398832\\
214	1.22737371398832\\
215	1.22737371398832\\
216	1.22737371398832\\
217	1.22737371398832\\
218	1.22737371398832\\
219	1.22737371398832\\
220	1.22737371398832\\
221	1.22737371398832\\
222	1.22737371398832\\
223	1.22737371398832\\
224	1.22737371398832\\
225	1.22737371398832\\
226	1.22737371398832\\
227	1.22737371398832\\
228	1.22737371398832\\
229	1.22737371398832\\
230	1.22737371398832\\
231	1.22737371398832\\
232	1.22737371398832\\
233	1.22737371398832\\
234	1.22737371398832\\
235	1.22737371398832\\
236	1.22737371398832\\
237	1.22737371398832\\
238	1.22737371398832\\
239	1.22737371398832\\
240	1.22737371398832\\
};
\addplot [color=mycolor1, dashdotted, line width=1.5pt, forget plot]
  table[row sep=crcr]{%
0	0\\
1	0\\
2	0\\
3	0\\
4	0\\
5	0\\
6	0\\
7	0\\
8	0\\
9	0\\
10	0\\
11	0\\
12	0\\
13	0\\
14	0\\
15	0\\
16	0\\
17	0\\
18	0\\
19	0\\
20	0\\
21	0\\
22	0\\
23	0\\
24	0\\
25	0\\
26	0\\
27	0\\
28	0\\
29	0\\
30	0\\
31	0\\
32	0\\
33	0\\
34	0\\
35	0\\
36	0\\
37	0\\
38	0\\
39	0\\
40	0\\
41	0\\
42	0\\
43	0\\
44	0\\
45	0.00694969400761563\\
46	0.128866112304087\\
47	0.128866112304087\\
48	0.128866112304087\\
49	0.128866112304087\\
50	0.128866112304087\\
51	0.128866112304087\\
52	0.128866112304087\\
53	0.128866112304087\\
54	0.128866112304087\\
55	0.128866112304087\\
56	0.128866112304087\\
57	0.128866112304087\\
58	0.128866112304087\\
59	0.128866112304087\\
60	0.128866112304087\\
61	0.144090578795124\\
62	0.144090578795124\\
63	0.144090578795124\\
64	0.144090578795124\\
65	0.144090578795124\\
66	0.144090578795124\\
67	0.144090578795124\\
68	0.144090578795124\\
69	0.144090578795124\\
70	0.144090578795124\\
71	0.144090578795124\\
72	0.144090578795124\\
73	0.144090578795124\\
74	0.144090578795124\\
75	0.144090578795124\\
76	0.144090578795124\\
77	0.144090578795124\\
78	0.144090578795124\\
79	0.144090578795124\\
80	0.144090578795124\\
81	1.50676208055807\\
82	1.50676208055807\\
83	1.50676208055807\\
84	1.50676208055807\\
85	1.50676208055807\\
86	1.50676208055807\\
87	1.50676208055807\\
88	1.50676208055807\\
89	1.50676208055807\\
90	1.50676208055807\\
91	1.50676208055807\\
92	1.50676208055807\\
93	1.50676208055807\\
94	1.50676208055807\\
95	1.50676208055807\\
96	1.50676208055807\\
97	1.50676208055807\\
98	1.50676208055807\\
99	1.50676208055807\\
100	1.50676208055807\\
101	1.50676208055807\\
102	1.50676208055807\\
103	1.50676208055807\\
104	1.50676208055807\\
105	1.50676208055807\\
106	1.50676208055807\\
107	1.50676208055807\\
108	1.50676208055807\\
109	1.50676208055807\\
110	1.50676208055807\\
111	1.6552251396466\\
112	1.6552251396466\\
113	1.6552251396466\\
114	1.6552251396466\\
115	1.6552251396466\\
116	1.6552251396466\\
117	1.6552251396466\\
118	1.6552251396466\\
119	1.6552251396466\\
120	1.6552251396466\\
121	1.6552251396466\\
122	1.6552251396466\\
123	1.6552251396466\\
124	1.6552251396466\\
125	1.6552251396466\\
126	1.6552251396466\\
127	1.6552251396466\\
128	1.6552251396466\\
129	1.6552251396466\\
130	1.6552251396466\\
131	1.6552251396466\\
132	1.6552251396466\\
133	1.6552251396466\\
134	1.6552251396466\\
135	1.6552251396466\\
136	1.6552251396466\\
137	1.6552251396466\\
138	1.6552251396466\\
139	1.6552251396466\\
140	1.6552251396466\\
141	1.6552251396466\\
142	1.6552251396466\\
143	1.6552251396466\\
144	1.6552251396466\\
145	1.6552251396466\\
146	1.6552251396466\\
147	1.6552251396466\\
148	1.6552251396466\\
149	1.6552251396466\\
150	1.6552251396466\\
151	1.6552251396466\\
152	1.6552251396466\\
153	1.6552251396466\\
154	1.6552251396466\\
155	1.6552251396466\\
156	1.6552251396466\\
157	1.6552251396466\\
158	1.6552251396466\\
159	1.6552251396466\\
160	1.6552251396466\\
161	1.6552251396466\\
162	1.6552251396466\\
163	1.6552251396466\\
164	1.6552251396466\\
165	1.6552251396466\\
166	1.6552251396466\\
167	1.6552251396466\\
168	1.6552251396466\\
169	1.6552251396466\\
170	1.6552251396466\\
171	1.6552251396466\\
172	1.6552251396466\\
173	1.6552251396466\\
174	1.6552251396466\\
175	1.6552251396466\\
176	1.6552251396466\\
177	1.6552251396466\\
178	1.6552251396466\\
179	1.6552251396466\\
180	1.6552251396466\\
181	1.6552251396466\\
182	1.6552251396466\\
183	1.6552251396466\\
184	1.6552251396466\\
185	1.6552251396466\\
186	1.6552251396466\\
187	1.6552251396466\\
188	1.6552251396466\\
189	1.6552251396466\\
190	1.6552251396466\\
191	1.6552251396466\\
192	1.6552251396466\\
193	1.6552251396466\\
194	1.6552251396466\\
195	1.6552251396466\\
196	1.6552251396466\\
197	1.6552251396466\\
198	1.6552251396466\\
199	1.6552251396466\\
200	1.6552251396466\\
201	1.6552251396466\\
202	1.6552251396466\\
203	1.6552251396466\\
204	1.6552251396466\\
205	1.6552251396466\\
206	1.6552251396466\\
207	1.6552251396466\\
208	1.6552251396466\\
209	1.6552251396466\\
210	1.6552251396466\\
211	1.6552251396466\\
212	1.6552251396466\\
213	1.6552251396466\\
214	1.6552251396466\\
215	1.6552251396466\\
216	1.6552251396466\\
217	1.6552251396466\\
218	1.6552251396466\\
219	1.6552251396466\\
220	1.6552251396466\\
221	1.6552251396466\\
222	1.6552251396466\\
223	1.6552251396466\\
224	1.6552251396466\\
225	1.6552251396466\\
226	1.6552251396466\\
227	1.6552251396466\\
228	1.6552251396466\\
229	1.6552251396466\\
230	1.6552251396466\\
231	1.6552251396466\\
232	1.6552251396466\\
233	1.6552251396466\\
234	1.6552251396466\\
235	1.6552251396466\\
236	1.6552251396466\\
237	1.6552251396466\\
238	1.6552251396466\\
239	1.6552251396466\\
240	1.6552251396466\\
};
\addplot [color=mycolor1, dashdotted, line width=1.5pt, forget plot]
  table[row sep=crcr]{%
0	0\\
1	0\\
2	0\\
3	0\\
4	0\\
5	0\\
6	0\\
7	0\\
8	0\\
9	0\\
10	0\\
11	0\\
12	0\\
13	0\\
14	0\\
15	0\\
16	0\\
17	0\\
18	0\\
19	0\\
20	0\\
21	0\\
22	0\\
23	0\\
24	0\\
25	0\\
26	0\\
27	0\\
28	0\\
29	0\\
30	0\\
31	0\\
32	0\\
33	0\\
34	0\\
35	0\\
36	0\\
37	0\\
38	0\\
39	0\\
40	0\\
41	0\\
42	0\\
43	0\\
44	0\\
45	0\\
46	0\\
47	0\\
48	0\\
49	0\\
50	0\\
51	0\\
52	0\\
53	0\\
54	0\\
55	0\\
56	0\\
57	0\\
58	0\\
59	0\\
60	0\\
61	0\\
62	0\\
63	0\\
64	0\\
65	0\\
66	0\\
67	0\\
68	0\\
69	0\\
70	0\\
71	0\\
72	0\\
73	0\\
74	0\\
75	0\\
76	0\\
77	0\\
78	0\\
79	0\\
80	0\\
81	0\\
82	0\\
83	0\\
84	0\\
85	0\\
86	0\\
87	0\\
88	0\\
89	0\\
90	0\\
91	0\\
92	0\\
93	0\\
94	0\\
95	0\\
96	0\\
97	0\\
98	0\\
99	0\\
100	0\\
101	0\\
102	0.0930547855278734\\
103	0.0930547855278734\\
104	0.130335932805527\\
105	0.144730662024402\\
106	0.144730662024402\\
107	0.161872872629927\\
108	0.161872872629927\\
109	0.322895412095348\\
110	0.322895412095348\\
111	2.08450819796106\\
112	2.08450819796106\\
113	2.08450819796106\\
114	2.08450819796106\\
115	2.08450819796106\\
116	2.08450819796106\\
117	2.08450819796106\\
118	2.08450819796106\\
119	2.08450819796106\\
120	2.08450819796106\\
121	2.08450819796106\\
122	2.08450819796106\\
123	2.08450819796106\\
124	2.08450819796106\\
125	2.08450819796106\\
126	2.08450819796106\\
127	2.08450819796106\\
128	2.08450819796106\\
129	2.08450819796106\\
130	2.08450819796106\\
131	2.08450819796106\\
132	2.08450819796106\\
133	2.08450819796106\\
134	2.08450819796106\\
135	2.08450819796106\\
136	2.08450819796106\\
137	2.08450819796106\\
138	2.08450819796106\\
139	2.08450819796106\\
140	2.08450819796106\\
141	2.08450819796106\\
142	2.08450819796106\\
143	2.08450819796106\\
144	2.08450819796106\\
145	2.08450819796106\\
146	2.08450819796106\\
147	2.08450819796106\\
148	2.08450819796106\\
149	2.08450819796106\\
150	2.08450819796106\\
151	2.08450819796106\\
152	2.08450819796106\\
153	2.08450819796106\\
154	2.08450819796106\\
155	2.08450819796106\\
156	2.08450819796106\\
157	2.08450819796106\\
158	2.08450819796106\\
159	2.08450819796106\\
160	2.08450819796106\\
161	2.08450819796106\\
162	2.08450819796106\\
163	2.08450819796106\\
164	2.08450819796106\\
165	2.08450819796106\\
166	2.08450819796106\\
167	2.08450819796106\\
168	2.08450819796106\\
169	2.08450819796106\\
170	2.08450819796106\\
171	2.08450819796106\\
172	2.08450819796106\\
173	2.08450819796106\\
174	2.08450819796106\\
175	2.08450819796106\\
176	2.08450819796106\\
177	2.08450819796106\\
178	2.08450819796106\\
179	2.08450819796106\\
180	2.08450819796106\\
181	2.08450819796106\\
182	2.08450819796106\\
183	2.08450819796106\\
184	2.08450819796106\\
185	2.08450819796106\\
186	2.08450819796106\\
187	2.08450819796106\\
188	2.08450819796106\\
189	2.08450819796106\\
190	2.08450819796106\\
191	2.08450819796106\\
192	2.08450819796106\\
193	2.08450819796106\\
194	2.08450819796106\\
195	2.08450819796106\\
196	2.08450819796106\\
197	2.08450819796106\\
198	2.08450819796106\\
199	2.08450819796106\\
200	2.08450819796106\\
201	2.08450819796106\\
202	2.08450819796106\\
203	2.08450819796106\\
204	2.08450819796106\\
205	2.08450819796106\\
206	2.08450819796106\\
207	2.08450819796106\\
208	2.08450819796106\\
209	2.08450819796106\\
210	2.08450819796106\\
211	2.08450819796106\\
212	2.08450819796106\\
213	2.08450819796106\\
214	2.08450819796106\\
215	2.08450819796106\\
216	2.08450819796106\\
217	2.08450819796106\\
218	2.08450819796106\\
219	2.08450819796106\\
220	2.08450819796106\\
221	2.08450819796106\\
222	2.08450819796106\\
223	2.08450819796106\\
224	2.08450819796106\\
225	2.08450819796106\\
226	2.08450819796106\\
227	2.08450819796106\\
228	2.08450819796106\\
229	2.08450819796106\\
230	2.08450819796106\\
231	2.08450819796106\\
232	2.08450819796106\\
233	2.08450819796106\\
234	2.08450819796106\\
235	2.08450819796106\\
236	2.08450819796106\\
237	2.08450819796106\\
238	2.08450819796106\\
239	2.08450819796106\\
240	2.08450819796106\\
};
\addplot [color=mycolor1, dashdotted, line width=1.5pt, forget plot]
  table[row sep=crcr]{%
0	0\\
1	0\\
2	0\\
3	0\\
4	0\\
5	0\\
6	0\\
7	0\\
8	0\\
9	0\\
10	0\\
11	0\\
12	0\\
13	0\\
14	0\\
15	0\\
16	0\\
17	0\\
18	0\\
19	0\\
20	0\\
21	0\\
22	0\\
23	0\\
24	0\\
25	0\\
26	0\\
27	0\\
28	0\\
29	0\\
30	0\\
31	0\\
32	0\\
33	0\\
34	0\\
35	0\\
36	0\\
37	0\\
38	0\\
39	0\\
40	0\\
41	0\\
42	0\\
43	0\\
44	0\\
45	0\\
46	0\\
47	0\\
48	0\\
49	0\\
50	0\\
51	0\\
52	0\\
53	0\\
54	0\\
55	0\\
56	0\\
57	0\\
58	0\\
59	0\\
60	0\\
61	0\\
62	0\\
63	0\\
64	0\\
65	0\\
66	0\\
67	0\\
68	0\\
69	0\\
70	0\\
71	0\\
72	0\\
73	0\\
74	0\\
75	0\\
76	0\\
77	0\\
78	0\\
79	0\\
80	0\\
81	0\\
82	0\\
83	0\\
84	0\\
85	0\\
86	0\\
87	0\\
88	0\\
89	0\\
90	0\\
91	0\\
92	0\\
93	0\\
94	0\\
95	0\\
96	0\\
97	0\\
98	0\\
99	0\\
100	0\\
101	0\\
102	0\\
103	0\\
104	0\\
105	0\\
106	0\\
107	0\\
108	0\\
109	0\\
110	0\\
111	0\\
112	0\\
113	0\\
114	0\\
115	0\\
116	0\\
117	0\\
118	0\\
119	0\\
120	0\\
121	0\\
122	0\\
123	0\\
124	0\\
125	0\\
126	0\\
127	0\\
128	0\\
129	0\\
130	0\\
131	0\\
132	0\\
133	0.00618802312921362\\
134	0.0102544206525418\\
135	0.148493209104263\\
136	0.148493209104263\\
137	0.148493209104263\\
138	0.148493209104263\\
139	0.148493209104263\\
140	0.148493209104263\\
141	0.148493209104263\\
142	0.148493209104263\\
143	0.148493209104263\\
144	0.148493209104263\\
145	0.148493209104263\\
146	0.148493209104263\\
147	0.148493209104263\\
148	0.148493209104263\\
149	0.148493209104263\\
150	2.69258990966353\\
151	2.69258990966353\\
152	2.69258990966353\\
153	2.69258990966353\\
154	2.69258990966353\\
155	2.69258990966353\\
156	2.69258990966353\\
157	2.69258990966353\\
158	2.69258990966353\\
159	2.69258990966353\\
160	2.69258990966353\\
161	2.69258990966353\\
162	2.69258990966353\\
163	2.69258990966353\\
164	2.69258990966353\\
165	2.69258990966353\\
166	2.69258990966353\\
167	2.69258990966353\\
168	2.69258990966353\\
169	2.69258990966353\\
170	2.69258990966353\\
171	2.69258990966353\\
172	2.69258990966353\\
173	2.69258990966353\\
174	2.69258990966353\\
175	2.69258990966353\\
176	2.69258990966353\\
177	2.69258990966353\\
178	2.69258990966353\\
179	2.69258990966353\\
180	2.69258990966353\\
181	2.69258990966353\\
182	2.69258990966353\\
183	2.69258990966353\\
184	2.69258990966353\\
185	2.69258990966353\\
186	2.69258990966353\\
187	2.69258990966353\\
188	2.69258990966353\\
189	2.69258990966353\\
190	2.69258990966353\\
191	2.69258990966353\\
192	2.69258990966353\\
193	2.69258990966353\\
194	2.69258990966353\\
195	2.69258990966353\\
196	2.69258990966353\\
197	2.69258990966353\\
198	2.69258990966353\\
199	2.69258990966353\\
200	2.69258990966353\\
201	2.69258990966353\\
202	2.69258990966353\\
203	2.69258990966353\\
204	2.69258990966353\\
205	2.69258990966353\\
206	2.69258990966353\\
207	2.69258990966353\\
208	2.69258990966353\\
209	2.69258990966353\\
210	2.69258990966353\\
211	2.69258990966353\\
212	2.69258990966353\\
213	2.69258990966353\\
214	2.69258990966353\\
215	2.69258990966353\\
216	2.69258990966353\\
217	2.69258990966353\\
218	2.69258990966353\\
219	2.69258990966353\\
220	2.69258990966353\\
221	2.69258990966353\\
222	2.69258990966353\\
223	2.69258990966353\\
224	2.69258990966353\\
225	2.69258990966353\\
226	2.69258990966353\\
227	2.69258990966353\\
228	2.69258990966353\\
229	2.69258990966353\\
230	2.69258990966353\\
231	2.69258990966353\\
232	2.69258990966353\\
233	2.69258990966353\\
234	2.69258990966353\\
235	2.69258990966353\\
236	2.69258990966353\\
237	2.69258990966353\\
238	2.69258990966353\\
239	2.69258990966353\\
240	2.69258990966353\\
};
\addplot [color=mycolor1, dashdotted, line width=1.5pt, forget plot]
  table[row sep=crcr]{%
0	0\\
1	0\\
2	0\\
3	0\\
4	0\\
5	0\\
6	0\\
7	0\\
8	0\\
9	0\\
10	0\\
11	0\\
12	0\\
13	0\\
14	0\\
15	0\\
16	0\\
17	0\\
18	0\\
19	0\\
20	0\\
21	0\\
22	0\\
23	0\\
24	0.101164432574991\\
25	0.996622394088004\\
26	1.27595055997904\\
27	2.4200886776039\\
28	2.4200886776039\\
29	2.4200886776039\\
30	2.4200886776039\\
31	2.4200886776039\\
32	2.4200886776039\\
33	2.4200886776039\\
34	2.4200886776039\\
35	2.4200886776039\\
36	2.4200886776039\\
37	2.4200886776039\\
38	2.4200886776039\\
39	2.4200886776039\\
40	2.4200886776039\\
41	2.4200886776039\\
42	2.4200886776039\\
43	2.4200886776039\\
44	2.4200886776039\\
45	2.4200886776039\\
46	2.4200886776039\\
47	2.4200886776039\\
48	2.4200886776039\\
49	2.4200886776039\\
50	2.4200886776039\\
51	2.4200886776039\\
52	2.4200886776039\\
53	2.4200886776039\\
54	2.4200886776039\\
55	2.4200886776039\\
56	2.4200886776039\\
57	2.4200886776039\\
58	2.4200886776039\\
59	2.4200886776039\\
60	2.4200886776039\\
61	2.4200886776039\\
62	2.4200886776039\\
63	2.4200886776039\\
64	2.4200886776039\\
65	2.4200886776039\\
66	2.4200886776039\\
67	2.4200886776039\\
68	2.4200886776039\\
69	2.4200886776039\\
70	2.4200886776039\\
71	2.4200886776039\\
72	2.4200886776039\\
73	2.4200886776039\\
74	2.4200886776039\\
75	2.4200886776039\\
76	2.4200886776039\\
77	2.4200886776039\\
78	2.4200886776039\\
79	2.4200886776039\\
80	2.4200886776039\\
81	2.4200886776039\\
82	2.4200886776039\\
83	2.4200886776039\\
84	2.4200886776039\\
85	2.4200886776039\\
86	2.4200886776039\\
87	2.4200886776039\\
88	2.4200886776039\\
89	2.4200886776039\\
90	2.4200886776039\\
91	2.4200886776039\\
92	2.4200886776039\\
93	2.4200886776039\\
94	2.4200886776039\\
95	2.4200886776039\\
96	2.4200886776039\\
97	2.4200886776039\\
98	2.4200886776039\\
99	2.4200886776039\\
100	2.4200886776039\\
101	2.4200886776039\\
102	2.4200886776039\\
103	2.4200886776039\\
104	2.4200886776039\\
105	2.4200886776039\\
106	2.4200886776039\\
107	2.4200886776039\\
108	2.4200886776039\\
109	2.4200886776039\\
110	2.4200886776039\\
111	2.4200886776039\\
112	2.4200886776039\\
113	2.4200886776039\\
114	2.4200886776039\\
115	2.4200886776039\\
116	2.4200886776039\\
117	2.4200886776039\\
118	2.4200886776039\\
119	2.4200886776039\\
120	2.4200886776039\\
121	2.4200886776039\\
122	2.4200886776039\\
123	2.4200886776039\\
124	2.4200886776039\\
125	2.4200886776039\\
126	2.4200886776039\\
127	2.4200886776039\\
128	2.4200886776039\\
129	2.4200886776039\\
130	2.4200886776039\\
131	2.4200886776039\\
132	2.4200886776039\\
133	2.4200886776039\\
134	2.4200886776039\\
135	2.4200886776039\\
136	2.4200886776039\\
137	2.4200886776039\\
138	2.4200886776039\\
139	2.4200886776039\\
140	2.4200886776039\\
141	2.4200886776039\\
142	2.4200886776039\\
143	2.4200886776039\\
144	2.4200886776039\\
145	2.4200886776039\\
146	2.4200886776039\\
147	2.4200886776039\\
148	2.4200886776039\\
149	2.4200886776039\\
150	2.4200886776039\\
151	2.4200886776039\\
152	2.4200886776039\\
153	2.4200886776039\\
154	2.4200886776039\\
155	2.4200886776039\\
156	2.4200886776039\\
157	2.4200886776039\\
158	2.4200886776039\\
159	2.4200886776039\\
160	2.4200886776039\\
161	2.4200886776039\\
162	2.4200886776039\\
163	2.4200886776039\\
164	2.4200886776039\\
165	2.4200886776039\\
166	2.4200886776039\\
167	2.4200886776039\\
168	2.4200886776039\\
169	2.4200886776039\\
170	2.4200886776039\\
171	2.4200886776039\\
172	2.4200886776039\\
173	2.4200886776039\\
174	2.4200886776039\\
175	2.4200886776039\\
176	2.4200886776039\\
177	2.4200886776039\\
178	2.4200886776039\\
179	2.4200886776039\\
180	2.4200886776039\\
181	2.4200886776039\\
182	2.4200886776039\\
183	2.4200886776039\\
184	2.4200886776039\\
185	2.4200886776039\\
186	2.4200886776039\\
187	2.4200886776039\\
188	2.4200886776039\\
189	2.4200886776039\\
190	2.4200886776039\\
191	2.4200886776039\\
192	2.4200886776039\\
193	2.4200886776039\\
194	2.4200886776039\\
195	2.4200886776039\\
196	2.4200886776039\\
197	2.4200886776039\\
198	2.4200886776039\\
199	2.4200886776039\\
200	2.4200886776039\\
201	2.4200886776039\\
202	2.4200886776039\\
203	2.4200886776039\\
204	2.4200886776039\\
205	2.4200886776039\\
206	2.4200886776039\\
207	2.4200886776039\\
208	2.4200886776039\\
209	2.4200886776039\\
210	2.4200886776039\\
211	2.4200886776039\\
212	2.4200886776039\\
213	2.4200886776039\\
214	2.4200886776039\\
215	2.4200886776039\\
216	2.4200886776039\\
217	2.4200886776039\\
218	2.4200886776039\\
219	2.4200886776039\\
220	2.4200886776039\\
221	2.4200886776039\\
222	2.4200886776039\\
223	2.4200886776039\\
224	2.4200886776039\\
225	2.4200886776039\\
226	2.4200886776039\\
227	2.4200886776039\\
228	2.4200886776039\\
229	2.4200886776039\\
230	2.4200886776039\\
231	2.4200886776039\\
232	2.4200886776039\\
233	2.4200886776039\\
234	2.4200886776039\\
235	2.4200886776039\\
236	2.4200886776039\\
237	2.4200886776039\\
238	2.4200886776039\\
239	2.4200886776039\\
240	2.4200886776039\\
};
\end{axis}
\end{tikzpicture}%

%% file: figures/K_sparsityx.tex
%
%
\definecolor{mycolor1}{rgb}{0.00000,0.44700,0.74100}%
\definecolor{mycolor2}{rgb}{0.85000,0.32500,0.09800}%
\definecolor{mycolor3}{rgb}{0.92900,0.69400,0.12500}%
\definecolor{mycolor4}{rgb}{0.49400,0.18400,0.55600}%
\begin{tikzpicture}

\begin{axis}[%
	width=0.96\figurewidth,
height=0.85\figurewidth,
at={(0\figurewidth, 0\figurewidth)},
scale only axis,
xmin=20,
xmax=240,
xlabel style={font=\color{white!15!black}},
xlabel={$t$ in $\mathtt{h}$},
ymin=0,
ymax=15,
ylabel style={font=\color{white!15!black},at={(axis description cs:0.05,0.5)}},
ylabel={$\bar{c}_{h}^{\mu_{\mathtt{max}}}(t)$},
xtick={20,50,100,150,200},
axis background/.style={fill=white},
legend style={at={(0.76,0.65)}, anchor=south west, legend cell align=left, align=left, draw=white!15!black}
]
\addplot [color=mycolor1, line width=1.5pt, mark=+, mark options={solid, mycolor1}, mark indices={1,2,3,91,92,151,152,10,20,30,40,50,60,70,80,90,100,110,120,130,140,150,160,170,180,190,200,210,220,230,240}]
table[row sep=crcr]{%
	0	0\\
	1	0\\
	2	0\\
	3	0\\
	4	0\\
	5	0\\
	6	0\\
	7	0\\
	8	0\\
	9	0\\
	10	0\\
	11	0\\
	12	0\\
	13	0\\
	14	0\\
	15	0\\
	16	0\\
	17	0\\
	18	0\\
	19	0\\
	20	0\\
	21	0\\
	22	0\\
	23	0\\
	24	0\\
	25	0\\
	26	0\\
	27	0\\
	28	0\\
	29	0\\
	30	0\\
	31	0\\
	32	0\\
	33	0\\
	34	0\\
	35	0\\
	36	0\\
	37	0\\
	38	0\\
	39	0\\
	40	0\\
	41	0\\
	42	0\\
	43	0\\
	44	0\\
	45	0\\
	46	0\\
	47	0\\
	48	0\\
	49	0\\
	50	0\\
	51	0\\
	52	0\\
	53	0\\
	54	0\\
	55	0\\
	56	0\\
	57	0\\
	58	0\\
	59	0\\
	60	0\\
	61	0\\
	62	0\\
	63	0\\
	64	0\\
	65	0\\
	66	0\\
	67	0\\
	68	0\\
	69	0\\
	70	0\\
	71	0\\
	72	0\\
	73	0\\
	74	0\\
	75	0\\
	76	0\\
	77	0\\
	78	0\\
	79	0\\
	80	0\\
	81	0\\
	82	0\\
	83	0\\
	84	0\\
	85	0\\
	86	0\\
	87	0\\
	88	0\\
	89	0\\
	90	0\\
	91	0\\
	92	0\\
	93	0\\
	94	0\\
	95	0\\
	96	0\\
	97	0\\
	98	0\\
	99	0\\
	100	0\\
	101	0\\
	102	0\\
	103	0\\
	104	0\\
	105	0\\
	106	0\\
	107	0\\
	108	0\\
	109	0\\
	110	0.0406007382100115\\
	111	0.178525581374781\\
	112	0.178525581374781\\
	113	0.178525581374781\\
	114	0.178525581374781\\
	115	0.178525581374781\\
	116	0.178525581374781\\
	117	0.178525581374781\\
	118	0.178525581374781\\
	119	0.178525581374781\\
	120	0.178525581374781\\
	121	0.178525581374781\\
	122	0.178525581374781\\
	123	0.178525581374781\\
	124	0.178525581374781\\
	125	0.178525581374781\\
	126	0.178525581374781\\
	127	0.178525581374781\\
	128	0.178525581374781\\
	129	0.178525581374781\\
	130	0.178525581374781\\
	131	0.178525581374781\\
	132	0.178525581374781\\
	133	0.178525581374781\\
	134	0.178525581374781\\
	135	0.178525581374781\\
	136	0.178525581374781\\
	137	0.178525581374781\\
	138	0.178525581374781\\
	139	0.178525581374781\\
	140	0.178525581374781\\
	141	0.178525581374781\\
	142	0.178525581374781\\
	143	0.178525581374781\\
	144	0.178525581374781\\
	145	0.178525581374781\\
	146	0.178525581374781\\
	147	0.178525581374781\\
	148	0.178525581374781\\
	149	0.178525581374781\\
	150	0.178525581374781\\
	151	0.178525581374781\\
	152	0.178525581374781\\
	153	0.178525581374781\\
	154	0.178525581374781\\
	155	0.178525581374781\\
	156	0.178525581374781\\
	157	0.178525581374781\\
	158	0.178525581374781\\
	159	0.178525581374781\\
	160	0.178525581374781\\
	161	0.178525581374781\\
	162	0.178525581374781\\
	163	0.178525581374781\\
	164	0.178525581374781\\
	165	0.178525581374781\\
	166	0.178525581374781\\
	167	0.178525581374781\\
	168	0.178525581374781\\
	169	0.178525581374781\\
	170	0.178525581374781\\
	171	2.57145858173366\\
	172	2.57145858173366\\
	173	2.57145858173366\\
	174	2.57145858173366\\
	175	2.57145858173366\\
	176	2.57145858173366\\
	177	2.57145858173366\\
	178	2.57145858173366\\
	179	2.57145858173366\\
	180	2.57145858173366\\
	181	2.57145858173366\\
	182	2.57145858173366\\
	183	2.57145858173366\\
	184	2.57145858173366\\
	185	2.57145858173366\\
	186	2.57145858173366\\
	187	2.57145858173366\\
	188	2.57145858173366\\
	189	2.57145858173366\\
	190	2.57145858173366\\
	191	2.57145858173366\\
	192	2.57145858173366\\
	193	2.57145858173366\\
	194	2.57145858173366\\
	195	2.57145858173366\\
	196	2.57145858173366\\
	197	2.57145858173366\\
	198	2.57145858173366\\
	199	2.57145858173366\\
	200	2.57145858173366\\
	201	2.57145858173366\\
	202	2.57145858173366\\
	203	2.57145858173366\\
	204	2.57145858173366\\
	205	2.57145858173366\\
	206	2.57145858173366\\
	207	2.57145858173366\\
	208	2.57145858173366\\
	209	2.57145858173366\\
	210	2.57145858173366\\
	211	2.57145858173366\\
	212	2.57145858173366\\
	213	2.57145858173366\\
	214	2.57145858173366\\
	215	2.57145858173366\\
	216	2.57145858173366\\
	217	2.57145858173366\\
	218	2.57145858173366\\
	219	2.57145858173366\\
	220	2.57145858173366\\
	221	2.57145858173366\\
	222	2.57145858173366\\
	223	2.57145858173366\\
	224	2.57145858173366\\
	225	2.57145858173366\\
	226	2.57145858173366\\
	227	2.57145858173366\\
	228	2.57145858173366\\
	229	2.57145858173366\\
	230	2.57145858173366\\
	231	2.57145858173366\\
	232	2.57145858173366\\
	233	2.57145858173366\\
	234	2.57145858173366\\
	235	2.57145858173366\\
	236	2.57145858173366\\
	237	2.57145858173366\\
	238	2.57145858173366\\
	239	2.57145858173366\\
	240	2.57145858173366\\
};
\addlegendentry{$K2$}

\addplot [color=mycolor2, line width=1.5pt, mark=x, mark options={solid, mycolor2}, mark indices={1,2,3,10,20,30,37,38,60,61,30,50,60,70,80,90,100,110,120,130,140,150,160,170,180,190,200,210,220,230,240}]
  table[row sep=crcr]{%
0	0\\
1	0\\
2	0\\
3	0\\
4	0\\
5	0\\
6	0\\
7	0\\
8	0\\
9	0\\
10	0\\
11	0\\
12	0\\
13	0\\
14	0\\
15	0\\
16	0\\
17	0\\
18	0\\
19	0\\
20	0\\
21	0\\
22	0\\
23	0\\
24	0\\
25	0\\
26	0\\
27	0\\
28	0\\
29	0\\
30	0\\
31	0\\
32	0\\
33	0\\
34	0\\
35	0\\
36	0\\
37	0\\
38	0\\
39	0\\
40	0\\
41	0\\
42	0\\
43	0\\
44	0\\
45	0\\
46	0\\
47	0\\
48	0\\
49	0\\
50	0\\
51	0\\
52	0\\
53	0\\
54	0\\
55	0\\
56	0.0258319499773168\\
57	0.1351911657352\\
58	0.1351911657352\\
59	0.1351911657352\\
60	0.1351911657352\\
61	0.1351911657352\\
62	0.1351911657352\\
63	0.1351911657352\\
64	0.1351911657352\\
65	0.1351911657352\\
66	0.1351911657352\\
67	0.1351911657352\\
68	0.1351911657352\\
69	0.1351911657352\\
70	0.1351911657352\\
71	0.1351911657352\\
72	0.1351911657352\\
73	0.1351911657352\\
74	0.1351911657352\\
75	0.1351911657352\\
76	0.1351911657352\\
77	0.1351911657352\\
78	0.1351911657352\\
79	0.1351911657352\\
80	3.02197958138586\\
81	3.02197958138586\\
82	3.02197958138586\\
83	3.02197958138586\\
84	3.02197958138586\\
85	3.02197958138586\\
86	3.02197958138586\\
87	3.02197958138586\\
88	3.02197958138586\\
89	3.02197958138586\\
90	3.02197958138586\\
91	3.02197958138586\\
92	3.02197958138586\\
93	3.02197958138586\\
94	3.02197958138586\\
95	3.02197958138586\\
96	3.02197958138586\\
97	3.02197958138586\\
98	3.02197958138586\\
99	3.02197958138586\\
100	3.02197958138586\\
101	3.02197958138586\\
102	3.02197958138586\\
103	3.02197958138586\\
104	3.02197958138586\\
105	3.02197958138586\\
106	3.02197958138586\\
107	3.02197958138586\\
108	3.02197958138586\\
109	3.02197958138586\\
110	3.02197958138586\\
111	3.02197958138586\\
112	3.02197958138586\\
113	3.02197958138586\\
114	3.02197958138586\\
115	3.02197958138586\\
116	3.02197958138586\\
117	3.02197958138586\\
118	3.02197958138586\\
119	3.02197958138586\\
120	3.02197958138586\\
121	3.02197958138586\\
122	3.02197958138586\\
123	3.02197958138586\\
124	3.02197958138586\\
125	3.02197958138586\\
126	3.02197958138586\\
127	3.02197958138586\\
128	3.02197958138586\\
129	3.02197958138586\\
130	3.02197958138586\\
131	3.02197958138586\\
132	3.02197958138586\\
133	3.02197958138586\\
134	3.02197958138586\\
135	3.02197958138586\\
136	3.02197958138586\\
137	3.02197958138586\\
138	3.02197958138586\\
139	3.02197958138586\\
140	3.02197958138586\\
141	3.02197958138586\\
142	3.02197958138586\\
143	3.02197958138586\\
144	3.02197958138586\\
145	3.02197958138586\\
146	3.02197958138586\\
147	3.02197958138586\\
148	3.02197958138586\\
149	3.02197958138586\\
150	3.02197958138586\\
151	3.02197958138586\\
152	3.02197958138586\\
153	3.02197958138586\\
154	3.02197958138586\\
155	3.02197958138586\\
156	3.02197958138586\\
157	3.02197958138586\\
158	3.02197958138586\\
159	3.02197958138586\\
160	3.02197958138586\\
161	3.02197958138586\\
162	3.02197958138586\\
163	3.02197958138586\\
164	3.02197958138586\\
165	3.02197958138586\\
166	3.02197958138586\\
167	3.02197958138586\\
168	3.02197958138586\\
169	3.02197958138586\\
170	3.02197958138586\\
171	3.02197958138586\\
172	3.02197958138586\\
173	3.02197958138586\\
174	3.02197958138586\\
175	3.02197958138586\\
176	3.02197958138586\\
177	3.02197958138586\\
178	3.02197958138586\\
179	3.02197958138586\\
180	3.02197958138586\\
181	3.02197958138586\\
182	3.02197958138586\\
183	3.02197958138586\\
184	3.02197958138586\\
185	3.02197958138586\\
186	3.02197958138586\\
187	3.02197958138586\\
188	3.02197958138586\\
189	3.02197958138586\\
190	3.02197958138586\\
191	3.02197958138586\\
192	3.02197958138586\\
193	3.02197958138586\\
194	3.02197958138586\\
195	3.02197958138586\\
196	3.02197958138586\\
197	3.02197958138586\\
198	3.02197958138586\\
199	3.02197958138586\\
200	3.02197958138586\\
201	3.02197958138586\\
202	3.02197958138586\\
203	3.02197958138586\\
204	3.02197958138586\\
205	3.02197958138586\\
206	3.02197958138586\\
207	3.02197958138586\\
208	3.02197958138586\\
209	3.02197958138586\\
210	3.02197958138586\\
211	3.02197958138586\\
212	3.02197958138586\\
213	3.02197958138586\\
214	3.02197958138586\\
215	3.02197958138586\\
216	3.02197958138586\\
217	3.02197958138586\\
218	3.02197958138586\\
219	3.02197958138586\\
220	3.02197958138586\\
221	3.02197958138586\\
222	3.02197958138586\\
223	3.02197958138586\\
224	3.02197958138586\\
225	3.02197958138586\\
226	3.02197958138586\\
227	3.02197958138586\\
228	3.02197958138586\\
229	3.02197958138586\\
230	3.02197958138586\\
231	3.02197958138586\\
232	3.02197958138586\\
233	3.02197958138586\\
234	3.02197958138586\\
235	3.02197958138586\\
236	3.02197958138586\\
237	3.02197958138586\\
238	3.02197958138586\\
239	3.02197958138586\\
240	3.02197958138586\\
};
\addlegendentry{$K4$}

\addplot [color=mycolor3, line width=1.5pt, mark=asterisk, mark options={solid, mycolor3}, mark indices={0,1,2,3,10,20,30,40,50,60,70,80,90,100,110,120,130,140,150,160,170,180,190,200,210,220,230,240}]
  table[row sep=crcr]{%
0	0\\
1	0\\
2	0\\
3	0\\
4	0\\
5	0\\
6	0\\
7	0\\
8	0\\
9	0\\
10	0\\
11	0\\
12	0\\
13	0\\
14	0\\
15	0\\
16	0\\
17	0\\
18	0\\
19	0\\
20	5.6558884133994\\
21	6.6689967520252\\
22	7.32747948088722\\
23	7.32747948088722\\
24	7.32747948088722\\
25	7.32747948088722\\
26	7.32747948088722\\
27	7.32747948088722\\
28	7.32747948088722\\
29	7.32747948088722\\
30	7.32747948088722\\
31	7.32747948088722\\
32	7.32747948088722\\
33	7.32747948088722\\
34	7.32747948088722\\
35	7.32747948088722\\
36	7.32747948088722\\
37	7.32747948088722\\
38	7.32747948088722\\
39	7.32747948088722\\
40	7.32747948088722\\
41	7.32747948088722\\
42	7.32747948088722\\
43	7.32747948088722\\
44	7.32747948088722\\
45	7.32747948088722\\
46	7.32747948088722\\
47	7.32747948088722\\
48	7.32747948088722\\
49	7.32747948088722\\
50	7.32747948088722\\
51	7.32747948088722\\
52	7.32747948088722\\
53	7.32747948088722\\
54	7.32747948088722\\
55	7.32747948088722\\
56	7.32747948088722\\
57	7.32747948088722\\
58	7.32747948088722\\
59	7.32747948088722\\
60	7.32747948088722\\
61	7.32747948088722\\
62	7.32747948088722\\
63	7.32747948088722\\
64	7.32747948088722\\
65	7.32747948088722\\
66	7.32747948088722\\
67	7.32747948088722\\
68	7.32747948088722\\
69	7.32747948088722\\
70	7.32747948088722\\
71	7.32747948088722\\
72	7.32747948088722\\
73	7.32747948088722\\
74	7.32747948088722\\
75	7.32747948088722\\
76	7.32747948088722\\
77	7.32747948088722\\
78	7.32747948088722\\
79	7.32747948088722\\
80	7.32747948088722\\
81	7.32747948088722\\
82	7.32747948088722\\
83	7.32747948088722\\
84	7.32747948088722\\
85	7.32747948088722\\
86	7.32747948088722\\
87	7.32747948088722\\
88	7.32747948088722\\
89	7.32747948088722\\
90	7.32747948088722\\
91	7.32747948088722\\
92	7.32747948088722\\
93	7.32747948088722\\
94	7.32747948088722\\
95	7.32747948088722\\
96	7.32747948088722\\
97	7.32747948088722\\
98	7.32747948088722\\
99	7.32747948088722\\
100	7.32747948088722\\
101	7.32747948088722\\
102	7.32747948088722\\
103	7.32747948088722\\
104	7.32747948088722\\
105	7.32747948088722\\
106	7.32747948088722\\
107	7.32747948088722\\
108	7.32747948088722\\
109	7.32747948088722\\
110	7.32747948088722\\
111	7.32747948088722\\
112	7.32747948088722\\
113	7.32747948088722\\
114	7.32747948088722\\
115	7.32747948088722\\
116	7.32747948088722\\
117	7.32747948088722\\
118	7.32747948088722\\
119	7.32747948088722\\
120	7.32747948088722\\
121	7.32747948088722\\
122	7.32747948088722\\
123	7.32747948088722\\
124	7.32747948088722\\
125	7.32747948088722\\
126	7.32747948088722\\
127	7.32747948088722\\
128	7.32747948088722\\
129	7.32747948088722\\
130	7.32747948088722\\
131	7.32747948088722\\
132	7.32747948088722\\
133	7.32747948088722\\
134	7.32747948088722\\
135	7.32747948088722\\
136	7.32747948088722\\
137	7.32747948088722\\
138	7.32747948088722\\
139	7.32747948088722\\
140	7.32747948088722\\
141	7.32747948088722\\
142	7.32747948088722\\
143	7.32747948088722\\
144	7.32747948088722\\
145	7.32747948088722\\
146	7.32747948088722\\
147	7.32747948088722\\
148	7.32747948088722\\
149	7.32747948088722\\
150	7.32747948088722\\
151	7.32747948088722\\
152	7.32747948088722\\
153	7.32747948088722\\
154	7.32747948088722\\
155	7.32747948088722\\
156	7.32747948088722\\
157	7.32747948088722\\
158	7.32747948088722\\
159	7.32747948088722\\
160	7.32747948088722\\
161	7.32747948088722\\
162	7.32747948088722\\
163	7.32747948088722\\
164	7.32747948088722\\
165	7.32747948088722\\
166	7.32747948088722\\
167	7.32747948088722\\
168	7.32747948088722\\
169	7.32747948088722\\
170	7.32747948088722\\
171	7.32747948088722\\
172	7.32747948088722\\
173	7.32747948088722\\
174	7.32747948088722\\
175	7.32747948088722\\
176	7.32747948088722\\
177	7.32747948088722\\
178	7.32747948088722\\
179	7.32747948088722\\
180	7.32747948088722\\
181	7.32747948088722\\
182	7.32747948088722\\
183	7.32747948088722\\
184	7.32747948088722\\
185	7.32747948088722\\
186	7.32747948088722\\
187	7.32747948088722\\
188	7.32747948088722\\
189	7.32747948088722\\
190	7.32747948088722\\
191	7.32747948088722\\
192	7.32747948088722\\
193	7.32747948088722\\
194	7.32747948088722\\
195	7.32747948088722\\
196	7.32747948088722\\
197	7.32747948088722\\
198	7.32747948088722\\
199	7.32747948088722\\
200	7.32747948088722\\
201	7.32747948088722\\
202	7.32747948088722\\
203	7.32747948088722\\
204	7.32747948088722\\
205	7.32747948088722\\
206	7.32747948088722\\
207	7.32747948088722\\
208	7.32747948088722\\
209	7.32747948088722\\
210	7.32747948088722\\
211	7.32747948088722\\
212	7.32747948088722\\
213	7.32747948088722\\
214	7.32747948088722\\
215	7.32747948088722\\
216	7.32747948088722\\
217	7.32747948088722\\
218	7.32747948088722\\
219	7.32747948088722\\
220	7.32747948088722\\
221	7.32747948088722\\
222	7.32747948088722\\
223	7.32747948088722\\
224	7.32747948088722\\
225	7.32747948088722\\
226	7.32747948088722\\
227	7.32747948088722\\
228	7.32747948088722\\
229	7.32747948088722\\
230	7.32747948088722\\
231	7.32747948088722\\
232	7.32747948088722\\
233	7.32747948088722\\
234	7.32747948088722\\
235	7.32747948088722\\
236	7.32747948088722\\
237	7.32747948088722\\
238	7.32747948088722\\
239	7.32747948088722\\
240	7.32747948088722\\
};
\addlegendentry{$K10$}

\addplot [color=mycolor4, line width=1.5pt, mark=square, mark options={solid, mycolor4}, mark indices={1,2,3,4,5,6,10,20,30,40,50,60,70,80,90,100,110,120,130,140,150,160,170,180,190,200,210,220,230,240}]
  table[row sep=crcr]{%
0	0\\
1	0\\
2	0\\
3	0\\
4	0\\
5	0\\
6	0\\
7	0\\
8	0\\
9	0\\
10	0\\
11	0\\
12	0\\
13	0\\
14	0\\
15	0\\
16	0\\
17	0\\
18	0\\
19	0\\
20	0\\
21	0\\
22	9.10708500114771\\
23	10.3639090656879\\
24	10.3639090656879\\
25	14.6127231758693\\
26	14.6127231758693\\
27	14.6127231758693\\
28	14.6127231758693\\
29	14.6127231758693\\
30	14.6127231758693\\
31	14.6127231758693\\
32	14.6127231758693\\
33	14.6127231758693\\
34	14.6127231758693\\
35	14.6127231758693\\
36	14.6127231758693\\
37	14.6127231758693\\
38	14.6127231758693\\
39	14.6127231758693\\
40	14.6127231758693\\
41	14.6127231758693\\
42	14.6127231758693\\
43	14.6127231758693\\
44	14.6127231758693\\
45	14.6127231758693\\
46	14.6127231758693\\
47	14.6127231758693\\
48	14.6127231758693\\
49	14.6127231758693\\
50	14.6127231758693\\
51	14.6127231758693\\
52	14.6127231758693\\
53	14.6127231758693\\
54	14.6127231758693\\
55	14.6127231758693\\
56	14.6127231758693\\
57	14.6127231758693\\
58	14.6127231758693\\
59	14.6127231758693\\
60	14.6127231758693\\
61	14.6127231758693\\
62	14.6127231758693\\
63	14.6127231758693\\
64	14.6127231758693\\
65	14.6127231758693\\
66	14.6127231758693\\
67	14.6127231758693\\
68	14.6127231758693\\
69	14.6127231758693\\
70	14.6127231758693\\
71	14.6127231758693\\
72	14.6127231758693\\
73	14.6127231758693\\
74	14.6127231758693\\
75	14.6127231758693\\
76	14.6127231758693\\
77	14.6127231758693\\
78	14.6127231758693\\
79	14.6127231758693\\
80	14.6127231758693\\
81	14.6127231758693\\
82	14.6127231758693\\
83	14.6127231758693\\
84	14.6127231758693\\
85	14.6127231758693\\
86	14.6127231758693\\
87	14.6127231758693\\
88	14.6127231758693\\
89	14.6127231758693\\
90	14.6127231758693\\
91	14.6127231758693\\
92	14.6127231758693\\
93	14.6127231758693\\
94	14.6127231758693\\
95	14.6127231758693\\
96	14.6127231758693\\
97	14.6127231758693\\
98	14.6127231758693\\
99	14.6127231758693\\
100	14.6127231758693\\
101	14.6127231758693\\
102	14.6127231758693\\
103	14.6127231758693\\
104	14.6127231758693\\
105	14.6127231758693\\
106	14.6127231758693\\
107	14.6127231758693\\
108	14.6127231758693\\
109	14.6127231758693\\
110	14.6127231758693\\
111	14.6127231758693\\
112	14.6127231758693\\
113	14.6127231758693\\
114	14.6127231758693\\
115	14.6127231758693\\
116	14.6127231758693\\
117	14.6127231758693\\
118	14.6127231758693\\
119	14.6127231758693\\
120	14.6127231758693\\
121	14.6127231758693\\
122	14.6127231758693\\
123	14.6127231758693\\
124	14.6127231758693\\
125	14.6127231758693\\
126	14.6127231758693\\
127	14.6127231758693\\
128	14.6127231758693\\
129	14.6127231758693\\
130	14.6127231758693\\
131	14.6127231758693\\
132	14.6127231758693\\
133	14.6127231758693\\
134	14.6127231758693\\
135	14.6127231758693\\
136	14.6127231758693\\
137	14.6127231758693\\
138	14.6127231758693\\
139	14.6127231758693\\
140	14.6127231758693\\
141	14.6127231758693\\
142	14.6127231758693\\
143	14.6127231758693\\
144	14.6127231758693\\
145	14.6127231758693\\
146	14.6127231758693\\
147	14.6127231758693\\
148	14.6127231758693\\
149	14.6127231758693\\
150	14.6127231758693\\
151	14.6127231758693\\
152	14.6127231758693\\
153	14.6127231758693\\
154	14.6127231758693\\
155	14.6127231758693\\
156	14.6127231758693\\
157	14.6127231758693\\
158	14.6127231758693\\
159	14.6127231758693\\
160	14.6127231758693\\
161	14.6127231758693\\
162	14.6127231758693\\
163	14.6127231758693\\
164	14.6127231758693\\
165	14.6127231758693\\
166	14.6127231758693\\
167	14.6127231758693\\
168	14.6127231758693\\
169	14.6127231758693\\
170	14.6127231758693\\
171	14.6127231758693\\
172	14.6127231758693\\
173	14.6127231758693\\
174	14.6127231758693\\
175	14.6127231758693\\
176	14.6127231758693\\
177	14.6127231758693\\
178	14.6127231758693\\
179	14.6127231758693\\
180	14.6127231758693\\
181	14.6127231758693\\
182	14.6127231758693\\
183	14.6127231758693\\
184	14.6127231758693\\
185	14.6127231758693\\
186	14.6127231758693\\
187	14.6127231758693\\
188	14.6127231758693\\
189	14.6127231758693\\
190	14.6127231758693\\
191	14.6127231758693\\
192	14.6127231758693\\
193	14.6127231758693\\
194	14.6127231758693\\
195	14.6127231758693\\
196	14.6127231758693\\
197	14.6127231758693\\
198	14.6127231758693\\
199	14.6127231758693\\
200	14.6127231758693\\
201	14.6127231758693\\
202	14.6127231758693\\
203	14.6127231758693\\
204	14.6127231758693\\
205	14.6127231758693\\
206	14.6127231758693\\
207	14.6127231758693\\
208	14.6127231758693\\
209	14.6127231758693\\
210	14.6127231758693\\
211	14.6127231758693\\
212	14.6127231758693\\
213	14.6127231758693\\
214	14.6127231758693\\
215	14.6127231758693\\
216	14.6127231758693\\
217	14.6127231758693\\
218	14.6127231758693\\
219	14.6127231758693\\
220	14.6127231758693\\
221	14.6127231758693\\
222	14.6127231758693\\
223	14.6127231758693\\
224	14.6127231758693\\
225	14.6127231758693\\
226	14.6127231758693\\
227	14.6127231758693\\
228	14.6127231758693\\
229	14.6127231758693\\
230	14.6127231758693\\
231	14.6127231758693\\
232	14.6127231758693\\
233	14.6127231758693\\
234	14.6127231758693\\
235	14.6127231758693\\
236	14.6127231758693\\
237	14.6127231758693\\
238	14.6127231758693\\
239	14.6127231758693\\
240	14.6127231758693\\
};
\addlegendentry{$K14$}

\end{axis}

\end{tikzpicture}%